\documentclass[11pt]{article}
\usepackage{amsfonts}
\usepackage{amsmath, amsthm, amssymb, bbm, bm, mathtools,thmtools}
\usepackage{float,xcolor,hyperref,multirow,graphicx,natbib}
\usepackage{pict2e,picture} 
\usepackage{tikz,pgfplots,tkz-euclide}   
\usetikzlibrary{automata, arrows.meta, positioning}
\usepackage{textcomp,comment}
\usepackage[flushleft]{threeparttable} 
\pgfplotsset{compat=1.10}
\usepackage[toc,page]{appendix}
\usepackage{booktabs}
\usepackage{physics}
\usepackage{makecell}
\usepackage{subcaption}

\newtheorem{assumption}{Assumption}
\newtheorem{theorem}{Theorem}
\newtheorem{example}{Example}
\newtheorem{remark}{Remark}
\newtheorem{proposition}{Proposition}

\newtheorem{lemma}[theorem]{Lemma}

\renewcommand\thmcontinues[1]{Continued}

\usepackage{chngcntr}
\counterwithin{example}{section}
\counterwithin{assumption}{section}

\newcommand{\argmax}{\operatorname*{argmax}}
\newcommand{\argmin}{\operatorname*{argmin}}

\newcommand{\E}{\mathbb{E}}

\begin{document}

 \title{Orthogonal Policy Learning Under Ambiguity\footnote{I am grateful to my advisors Martin Weidner and Toru Kitagawa for their help and support throughout my graduate studies. I would also like to thank Timothy Christensen, Sukjin Han, Whitney Newey, Tomasz Olma, Joris Pinkse, Andrei Zeleneev, as well as participants to the NeurIPS EconML Workshop, IAAE 2022 Conference, Bristol Econometric Study Group for helpful comments and discussions. \newline$^\dagger$Department of Economics, University College London. E-mail: \href{mailto:uctpdad@ucl.ac.uk}{\texttt{uctpdad@ucl.ac.uk}}}}\author{Riccardo D'Adamo$^\dagger$}
 
  \maketitle
  \begin{center}
  Job Market Paper\\
  \href{https://rdadamo.github.io/mywebsite/JMP_Dadamo.pdf}{[{\color{blue} Click here for the latest version}]}
\end{center}   

\begin{abstract}
This paper studies the problem of estimating individualized treatment rules when treatment effects are partially identified, as it is often the case with observational data. 
By drawing connections between the treatment assignment problem and classical decision theory, we characterize several notions of optimal treatment policies in the presence of partial identification. The proposed framework allows to incorporate user-defined constraints on the policies, such as restrictions for transparency or interpretability, while also ensuring computational feasibility.
We show that partial identification leads to a novel statistical learning problem with risk directionally -- but not fully -- differentiable with respect to an infinite-dimensional nuisance component. We propose an estimation procedure that ensures Neyman-orthogonality with respect to the nuisance component and provide statistical guarantees that depend on the amount of concentration around the points of non-differentiability in the data-generating process. The proposed method is illustrated using data from the Job Partnership Training Act study.  


\end{abstract}
\newpage
\section{Introduction}
The problem of choosing an optimal treatment assignment based on data is ubiquitous in economics and other fields, including medicine and marketing. Individuals often display heterogeneous responses to the same treatment. Decision-makers in policy and industry are therefore interested in leveraging the growing availability of rich granular data to tailor treatment assignment to individuals based on their characteristics. As a result, a fast-growing literature has emerged focused on developing procedures for estimation of individualized treatment rules. While a variety of approaches have been recently established, these typically assume that the available data allow to provide credible point estimates for the effect of the treatment, that is treatment effects are point identified. While of important stylized value, this assumption is often hard to justify in many empirical settings. For example, economists have long been aware that popular quasi-experimental and observational research designs, such as instrumental variables (IV), allow to point identify treatment effects only for specific sub-populations \citep{imbensangrist1994}. Even in randomized control trials, point identification of the treatment effects is often precluded due to non-random attrition, e.g.\@ when participants dropout from a program or the researcher is denied information on the outcome variable \citep{lee2009}. In such settings, the data may only provide partial knowledge about the treatment response in the form of credible bounds, i.e.\@ the treatment effects are \textit{partially identified}. As a result, the decision-maker may have ambiguous evidence on whether a candidate policy should be preferred to another, so that only a partial ordering of policies can be deduced in general.
While informative from a scientific perspective, a partial ordering of policies is unsatisfying when the ultimate goal of the analysis is to select a single policy to be implemented in the real world.
In this scenario, a decision-maker has to confront two sources of ambiguity. The first source concerns ambiguous knowledge of the treatment response $\tau$ conditional on knowledge of distribution of the data $P$, due to partial identification. The second source is the lack of knowledge of the distribution $P$, which must be estimated from the data.


In this paper, we develop methods to handle both sources of ambiguity within the framework of ``empirical welfare maximization" \citep{kitagawatetenov2018}, also referred to as ``policy learning" \citep{atheywager2021}. This approach considers treatment policies that are exogenously constrained to have low complexity in terms of Vapnik-Chervonenkis (VC) dimension. This encompasses many practical settings of interest, as policies often have to satisfy requirements imposed for institutional or practical reasons, such as fairness, budget or interpretability. The empirical welfare maximization (EWM) method selects the optimal policy as the maximizer of the empirical analogue of the population welfare, formulated as the average of the individual outcomes in the target population. The EWM estimation procedure has the convenient structure of an empirical risk minimization problem, which is exploited by \citet{kitagawatetenov2018} and \citet{atheywager2021} to study its statistical properties.

We extend the EWM framework to settings with partial identification by making several contributions. First, we study the problem of assigning treatment under partial identification at the population level (i.e.\@ where the distribution of the data $P$ is known) from a general perspective. In particular, we show how classic optimality criteria for decision under ambiguity, such as minimax risk and minimax regret, can be applied in the context of welfare maximization. Our unified framework accommodates different attitudes towards ambiguity and a wide range of popular identification assumptions, including \cite{manski1990} and \cite{manskiPepper} bounds. Our analysis delivers several notions of optimal treatment policies, which we refer to as \textit{ambiguity-robust}: they are ``robust" in the sense that each of them delivers a notion of single optimal policy in the presence of partial identification, while they all reduce to the same optimal treatment assignment in the special case of point-identification. As part of this analysis, we establish general conditions on the identification sets under which the treatment assignment problem can be expressed in a simplified form, leading to computationally tractable sample analogues. In particular, we show that all ambiguity-robust policies can be represented as maximizers of a ``surrogate" welfare, in which identification bounds are combined to form a proxy for the partially identified CATE.  The surrogate welfare depends on several nuisance components, and its specific form is determined by the identification assumptions and attitude towards ambiguity held by the decision-maker.
 
We then propose an algorithm for computing the estimated ambiguity-robust policy and provide statistical guarantees on its performance in terms of the regret convergence of the surrogate welfare. Similarly to \citet{atheywager2021} and \cite{fosterSyrgkanis}, our procedure leverages insights from the literature on double/de-biased machine learning \citep{chernozhukov2018} by making use of Neyman-orthogonalized estimates of the surrogate welfare. This, coupled with sample-splitting, allows us to guarantee 
 fast rates of convergence for the estimated ambiguity-robust optimal policy while imposing minimal requirements on the estimation of the nuisance components.
One unique feature of the partially identified setting studied in this paper is the restricted degree of smoothness enjoyed by the welfare criterion. In particular, we show that popular choices of identification assumptions and optimality criteria for choice under ambiguity lead to surrogate welfare criteria that are only \textit{directionally differentiable} with respect to the data-generating process. We highlight the importance of this feature for the problem at hand and develop new theoretical results showing how the extent of non-differentiability in the data-generating process affects the statistical properties of the learning procedure. To the best of our knowledge, we are the first to investigate the role of non-differentiabilities in the context of semiparametric statistical learning problems. Our results are therefore of independent interest and may be relevant beyond the treatment assignment problem of this paper. 

Finally, we apply the proposed method to experimental data from the Job Training Partnership Act study, a dataset that has been extensively used to study the effect of subsidized job training on labor market outcomes. We study the optimal participation of workers into the job training programme based on their education and previous earnings, and show that allowing for partial identification delivers substantially different programme participation policies compared to existing methods that assume point-identification.

\subsection{Related literature}

The results of this paper contribute to the recent literature on EWM methods, e.g.\@ \citet{kitagawatetenov2018}, \citet{atheywager2021}, \citet{mbakopTabord}, \cite{viviano2021}, \citet{sun2021}, and more broadly to the literature studying statistical treatment choice, including \cite{manski2004}, \cite{dehejia2005},  \cite{hiranoPorter2009ECMA}, \cite{stoye2009}, \cite{chamberlain2011}, \cite{christensenMoonSchor2020}, \cite{kitagawaLeeQiu2022}.\footnote{See also \cite{hiranoPorter2020HoE} and referenes therein.}

\citet{kitagawatetenov2018} introduced the EWM method and provided theoretical results showing its optimality when implemented with experimental data. \citet{atheywager2021} leverage insights from the recent literature on orthogonal machine learning \citep{chernozhukov2018} and propose doubly-robust estimation of the treatment effect 
which leads to optimal learning rates even with observational data. We build on their work by adopting Neyman-orthogonal estimates while we relax the fundamental assumption that treatment effects are point identified. \citet{cuiTchetgen2021} also develop procedures for learning optimal treatments rules with instrumental variables but consider unconstrained policy classes. Similarly to \cite{atheywager2021}, they ensure point-identification of treatment response by restricting their analysis to the effect on compliers.

\citet{kasy2016}, \cite{han2019} and \citet{Byambadalai2022} provide methods for comparing policies in the presence of covariates and partial identification of treatment effects. The focus of their work is on characterizing the partial ordering of policies in terms of  their associated welfare rather than resolving the ambiguity and estimating an optimal treatment rule.

In a series of papers, \citeauthor{manski2009} (\citeyear{manski2009}, \citeyear{manski2010}, \citeyear{manski2011}) studies the problem of a social planner who must choose treatment for a population under partial knowledge of the treatment response in the absence of covariates. He shows that when the sign of the treatment effect is ambiguous, the minimax regret criterion leads to policies that randomize treatment in the population. While our study of the population problem is inspired by Manski's work in this area, the focus of our paper is on deterministic rules assigning individualized treatment, i.e.\@ based on (potentially continuous) covariates. \cite{STOYE2012138}, \cite{ishiharaKitagawa} and \cite{yata} consider treatment assignment under partial identification from a finite-sample minimax perspective, while \cite{christensenMoonSchor2020} adopt a local-asymptotic approach. However, these works do not consider individualization of the treatment assignment.

More closely related to our work is \citet{kallus2018}, who extend the EWM framework to learn an optimal policy in the presence of partially identified treatment effects under violations of unconfoundedness. In particular, they target welfare improvement with respect to a baseline pre-existing policy and consider partial-identification of the welfare criterion within Rosenbaum's sensitivity model \citep{rosenbaum1987}. \cite{christensenAdjaho} and \cite{kido2022} examine policies with maximin welfare guarantees when the target population lies in a Wasserstein neighborhood of the experimental population. The identification assumptions (and associated estimation procedures) considered in these papers are distinct and do not nest those covered by our framework. As a result, our contributions are complementary to these works. 

\citet{russell2020} considers estimation of the optimal policy under partial identification within a ``probably approximately correct" learning framework \citep{valiant}. His proposed procedure has the advantage of side-stepping direct estimation of the identified set, and can be applied in the context of incomplete models for which the identification bounds cannot typically be obtained in closed form. However, in the context of the identification assumptions considered in this paper (e.g.\@ Manski bounds), the theoretical results in \citet{russell2020} require that the covariates have discrete support. On the other hand, our proposed procedure requires computation of the identification bounds in closed form but accommodates continuous covariates.

In independent work, \citet{Pu_2021} study policy learning under ambiguity from a classification perspective and derive an optimal policy which coincides with one notion of ambiguity-robust policy studied in this paper. However, our estimation procedure crucially differs from theirs for the use of Neyman-orthogonalization which, combined with a refined proof-strategy that accounts for the lack of full-differentiability in the welfare criterion, allows us to guarantee considerably faster rates of convergence. In this sense, our results extend and improve on those in \citet{Pu_2021}.

Finally, we contribute to a body of literature dealing with estimation and inference for directionally-differentiable functionals. \citet{hiranoPorter2012ECMA} show that if a target estimand is not differentiable in the parameters of the data distributions, then no asymptotically unbiased or regular estimator exists. \cite{ponomarev} studies efficient estimation of directionally differentiable functionals from a local minimax perspective. \citet{fangSantos2019} and \citet{kitagawaEtAl2020} provide inference results for directionally differentiable functions from a frequentist and Bayesian perspective, respectively. 
Also motivated by partial identification, \cite{christensenMoonSchor2020} consider estimation of treatment rules when the welfare criterion is only directionally differentiable with respect to a finite-dimensional nuisance component. Our framework instead involves infinite-dimensional nuisance components and therefore our  analysis must account for the lack of differentiability with novel theoretical results that substantially differ from \cite{christensenMoonSchor2020}.

The rest of this paper is organized as follows. Section \ref{sec:Setup} introduces the setup. Section \ref{sec:policyDerivation} presents several notions of ambiguity-robust optimal policies. Section \ref{sec:Estimation} presents the proposed estimation procedure for the ambiguity-robust optimal policy. Section \ref{sec:Theory} provides statistical guarantees for the estimated optimal policy. Section \ref{sec:empiricalApplication} presents an empirical illustration based on the Job Training Partnership Act Study. Section \ref{sec:Conclusion} concludes the paper. Proofs and extensions are given in the Appendix.



~\\
\noindent\textbf{Notation}. Throughout the paper, for $d\in \mathbb{N}$, let $\mathbb{R}^d$ denote the Euclidean space, with $\|\cdot \|_p$ and $\langle \cdot, \cdot \rangle$ being the usual $\ell_p$-norm and inner product, respectively. For two vectors $x \in \mathbb{R}^p$ and $y \in \mathbb{R}^q$, $x\subset y$ means that $x$ is a sub-vector of $y$. For a symmetric matrix  $A$, $\lambda_{{\rm max}}(A)$ denotes its largest eigenvalue. Unless otherwise stated, the expectation $\mathbb{E}[\cdot ]$, probability $\mathbb{P}(\cdot)$, and variance ${\rm Var}(\cdot)$ operators will be taken with respect to the underlying distribution of observables $P$. Given a random variable $Z\in \mathcal{Z}$ with $\mathcal{Z}\subseteq \mathbb{R}^d$, the associated probability measure $P_Z$, and a function $f\,:\, \mathcal{Z} \to \mathcal{W}$ with $\mathcal{W}\subseteq\mathbb{R}^q$, we define $\norm{f}_{Lp(P_Z)} = \left( \mathbb{E}_{P_Z} \left[ \|f(Z)\|^p_p \right]\right)^{1/p}$ for $p\in(0\,\infty)$. We extend this definition to $p=\infty$ in the natural way. For a sequence of real numbers $x_n$ and $y_n$, $x_n=o(y_n)$ and $x_n=O(y_n)$ mean, respectively, that $x_n/y_n\to 0$ and $x_n\leq C y_n$ for some constant $C$ as $n\to \infty$. For real numbers $a,b$, $a\lesssim b$ means that there exists a constant $C$ such that $a\leq C b$. For a positive real number $a$, $\lfloor a \rfloor$ denotes its nearest smallest integer.  The notation  $\rightarrow_p$ denotes convergence in probability. 

\section{Setup}\label{sec:Setup}

	Let $Y_i\in \mathbb{R}$ be an outcome measuring utility, $D_i\in\{0,1\}$ a binary treatment, $X_i \in \mathcal{X}\subseteq\mathbb{R}^{k_x}$ a set of pre-treatment covariates for an individual $i$ from an i.i.d.\@ population  of interest. We use standard notation to define the potential outcomes $Y_i(0), Y_i(1)$. The conditional average treatment effect (CATE) $\tau: \mathcal{X}\to \mathbb{R}$ is then defined as \begin{align*}
		\tau(x)= y_1(x)-y_0(x),\quad y_d(x)=\mathbb{E}[Y_i(d)|X_i=x ],\, d=0,1,
	\end{align*}
	where the expectation is taken with respect to the distribution of the population, and we will henceforth suppress the $i$-subscript for convenience. The decision-maker (DM) is interested in choosing a deterministic treatment assignment rule (or policy) $\pi:\mathcal{X}\to \{0,1\}$, which maps from the support of individual pre-treatment covariates to the binary decision ``treat" ($\pi(x)=1$) or ``do not treat" ($\pi(x)=0$). 
Following \citet{manski2004}, we define the utilitarian social welfare associated with a policy $\pi$ and a given configuration of the expected potential outcomes $y_0(\cdot),y_1(\cdot)$ as
	\begin{equation}
	\begin{aligned}
		W_{y_0,y_1}(\pi)&= \E_{P_X}\left[y_1(X)\cdot\pi(X) +y_0(X)\cdot(1-\pi(X))\right]\\
		&=\underbrace{\E_{P_X}\left[\pi(X)\cdot\tau(X)\right]}_{=:I_\tau(\pi)} + \E_{P_X}\left[y_0(X)\right],
	\end{aligned}
	\end{equation}
where $I_\tau(\pi)$ represents the average impact of policy $\pi$. The optimal policy for a given configuration of the CATE function is the one that maximizes the associated welfare:
	\begin{align}\label{generalOptimalPolicy}
		\pi^*= \argmax_{\pi \in \Pi }W_{y_0,y_1}(\pi)=\argmax_{\pi \in \Pi }I_{\tau}(\pi),
	\end{align}
where $\Pi$ is a family of candidate policies.\footnote{Throughout the paper, we will assume that the maximization problem in \eqref{generalOptimalPolicy} has at least one solution. If multiple solutions exist, the DM is assumed to arbitrarily pick $\pi^*$ from the set of maximizers.} The DM has knowledge of the CATE through the distribution $P\in \mathcal{P}$ of observable random variables $W$, where $(Y,D,X)\subseteq W$. In particular, we denote $\mathcal{T}(P)$ the set of plausible CATE functions associated with a certain distribution of observables. When the DM has perfect knowledge of $P$ and $\mathcal{T}(P)$ is a singleton, i.e.\@ $\tau$ is point-identified, she can obtain $\pi^*$ by solving \eqref{generalOptimalPolicy}.

Suppose now that $\mathcal{T}(P)$ is a non-singleton set, i.e.\@ $\tau$ is partially identified. In that case, even under perfect knowledge of $P$, there exists a set of plausible values for the impact $I_\tau(\pi)$ of a candidate policy $\pi$. Notice that partial identification of the CATE does not necessarily imply that the DM cannot obtain the optimal policy $\pi^*$. In particular, it is easy to see that under point-identification of the CATE one has $\pi^*(x)=\mathbbm{1}\{\tau(x)\geq 0\}$ when the class of candidate policies $\Pi$ is unrestricted, so that identification of the sign of the CATE is sufficient to obtain the optimal policy.\footnote{\cite{cuiTchetgen2021} study a case in which sole point-identification of the sign of the CATE via an instrumental variable allows to obtain the optimal policy.} However, the unrestricted policy class has limited relevance in many practical settings. For example, the policy space $\Pi$ may be exogenously constrained for institutional reasons, e.g.\@ as policies may be required to satisfy specific requirements for budget, fairness or interpretability. While the DM may still hope that his specification of $\Pi$ contains the first-best policy $\mathbbm{1}\{\tau(X)\geq 0\} $, it is useful to interpret $\pi^*$ as the ``best-in-class" policy for the chosen class $\Pi$, when this does not contain the first-best. When $\Pi$ is constrained, the DM is not able to obtain $\pi^*$ in general without full knowledge of the CATE, although a partial ordering of policies can still be deduced \citep[see][]{kasy2016,han2019,Byambadalai2022}.

Under partial identification, the DM therefore faces two sources of ambiguity. First, she does not know the distribution $P$. However, we assume that she has access to a random sample $(W_i)_{i=1,\dots,n}$ from which she can learn about $P$. Second, she does not have knowledge about $\tau$ within the identified-set $\mathcal{T}(P)$, even under perfect knowledge of $P$. The broad objective of this paper is to provide a framework that allows the DM to handle both sources of ambiguity. We will approach the problem in two steps. First, we will study the decision problem faced by the DM under perfect knowledge of $P$. In particular, we will handle the ambiguity arising from partial identification of $\tau$ using well-known optimality criteria for decision under ambiguity. Each of the optimality criteria we consider will deliver a corresponding notion of optimal policy, which we call ``ambiguity-robust". The ambiguity-robust optimal policy is a unique treatment assignment rule that is preferred to all other policies in $\Pi$ according to preferences of the DM, and that coincides with the usual notion of optimal policy $\pi^*$ in \eqref{generalOptimalPolicy} in the special case of point identification of the CATE. In the next section, we study several notions of ambiguity-robust optimal policy.

In the second part of our analysis, we study how to handle the ambiguity in $P$ by showing how the random sample $(W_i)_{i=1,\dots,n}$ can be used to obtain an estimate $\widehat \pi_n$ for the ambiguity-robust optimal policy. The estimation procedure and the associated statistical guarantees are presented in Section \ref{sec:Estimation} and \ref{sec:Theory}, respectively.


\begin{remark}
Unrestricted policy classes may also be precluded for practical reasons related to the estimation of the optimal policy.
For example, the researcher may need to condition on a large number of covariates $X$ for identification of the treatment effects, but only be interested in assigning treatment based on a restricted set of the covariates $\widetilde X \subset X$ (e.g.\@ because she may not observe the full set of covariates when assigning treatment to new individuals from the population). In that case, a practical way to side-step computation of an estimate for the lower-dimensional CATE, $\mathbb{E}[Y_i(1)|\widetilde{X}_i=\widetilde{x}]-\mathbb{E}[Y_i(0)|\widetilde{X}_i=\widetilde{x} ]$, is to impose restrictions directly on the policy class $\Pi$ and estimate the optimal policy based on the estimated higher-dimensional CATE via the sample analogue of \eqref{generalOptimalPolicy}.
\end{remark}

\section{Ambiguity-robust optimal policies}\label{sec:policyDerivation}
The study of decision under ambiguity has a long tradition in decision theory and has received considerable attention in the context of treatment assignment problems (see \citeauthor{manski2011}, \citeyear{manski2011}, for a review). In this section we review some classical optimality criteria for decision under ambiguity and study how they can be applied in the context of the treatment assignment problem at hand, leading to several notions of ambiguity-robust optimal policy.

A well-known optimality criterion for decision under ambiguity is minimax risk \citep[see, e.g.][]{wald_1950}. In the context of our treatment assignment problem we can interpret welfare as negative risk, and this criterion leads to the optimal \textit{maximin welfare} policy
\begin{align}\label{maximinWelfare}
	\pi^*_{\texttt{MMW}} = \argmax_{\pi \in \Pi } \min_{\left(y_0,y_1\right)\in \mathcal{Y}(P) }W_{y_0,y_1}(\pi),
\end{align}  
where $\mathcal{Y}(P)$ is the ambiguity set for $\left(y_0(\cdot),y_1(\cdot)\right)$ identified from the distribution $P$ of observables random variables. 
The optimal maximin welfare policy maximizes the lowest possible welfare under any configuration of the expected potential outcome functions in the identified set $\mathcal{Y}(P)$.
An alternative application of minimax risk optimality in the context of treatment assignment is \textit{maximin impact}, leading to the optimal policy
\begin{align}\label{maximinImpact}
	\pi^*_{\texttt{MMI}} = \argmax_{\pi \in \Pi } \min_{\tau \in \mathcal{T}(P) }I_\tau(\pi),
\end{align}
where $\mathcal{T}(P)$ denotes the ambiguity set for the CATE function. The optimal maximin impact policy maximizes the lowest possible impact under any configuration of the CATE in the identified set $\mathcal{T}(P)$. Notice that the minimax welfare criterion reflects an extreme degree of pessimism with regards to outcomes associated with both treatment and non-treatment scenarios; on the other hand, the minimax impact criterion reflects an extreme degree of pessimism with regards to the impact of the policy, thus directly raising the threshold for treatment.\footnote{In the empirical application of Section \ref{sec:empiricalApplication}, both minimax welfare and minimax impact criteria result in $\pi(x)=0$ for the entire population.} Despite its intuitive appeal, minimax optimality has been criticised for being too conservative and often delivering decisions that are especially sensitive to changes in the ambiguity set.\footnote{In his classic textbook, Berger goes as far as saying that ``In actually making decisions, the use of the minimax principle is definitely suspect." \citep{Berger}.} 

An alternative criterion that alleviates some of these concerns is \textit{minimax regret}, with corresponding optimal policy
\begin{equation}\label{BayesMinMaxRegret}
\begin{aligned}
	\pi^*_{\texttt{MMR}}&=\argmin_{\pi \in \Pi } \max_{\left(y_0,y_1\right) \in \mathcal{Y}(P) }\left[ \left(\max_{\pi\, : \, \mathcal{X}\to\{0,1\}}W_{y_0,y_1}(\pi)\right) - W_{y_0,y_1}(\pi)\right]\\
	&=\argmin_{\pi \in \Pi } \max_{\tau \in \mathcal{T}(P) }\left[ \left(\max_{\pi\, : \, \mathcal{X}\to\{0,1\}}I_\tau(\pi)\right) - I_\tau(\pi)\right],
\end{aligned}
\end{equation}
The minimax regret criterion delivers a policy that minimizes the largest possible distance between attained welfare and the highest level of welfare attainable by the ``oracle" treatment rule $\pi^*=\mathbbm{I}\left\{\tau(x)\geq0\right\}$ that has knowledge of the true $\tau$. Minimax regret optimality has been advocated by \citet{manski2004} for its balanced consideration of the possible states of nature and for delivering more ``reasonable" decisions rules in practice, compared to minimax risk approaches.
\begin{remark}
An alternative version of the minimax regret criterion is minimax regret with respect to the welfare attained by the best-in-class policy in $\Pi$, resulting in the objective
\begin{align}\label{MinMaxRegret}
	\pi^*_{\texttt{MMR2}} = \argmin_{\pi \in \Pi } \max_{\tau \in \mathcal{T}(P) }\left[\left( \max_{\pi \in \Pi}I_\tau(\pi)\right) - I_\tau(\pi)\right].
\end{align}
While these two versions of the minimax regret criterion can be expected to enjoy similar properties, the first version we have considered is considerably more tractable. In fact, the innermost maximization in (\ref{BayesMinMaxRegret}) has the closed-form solution $\max_{\pi\, : \, \mathcal{X}\to\{0,1\}}W_\tau(\pi)= \E_{P_X}\left[\max\left\{\tau(X),0\right\}\right]$. As we show in Proposition \ref{prop:minimaxRegret} below, this allows to more explicitly characterize the properties of the optimization problem and the resulting optimal policy, as well as reduce the computational burden in solving the empirical analogue of the problem. For this reason we will focus on the version in (\ref{BayesMinMaxRegret}) of the criterion. We also note that whenever the class $\Pi$ is ``well-specified", in the sense that $\mathbbm{I}\left\{\tau(x)\geq0\right\}\in \Pi$ for all $\tau\in\mathcal{T}(P)$, the two optimality criteria are equivalent.
\end{remark}
One critical drawback in the application of the optimality criteria just presented to the treatment assignment problem  of this paper is that the optimal policies cannot be obtained in closed form. This is due to the form of \eqref{maximinWelfare}, \eqref{maximinImpact} and (\ref{BayesMinMaxRegret}) involving several nested optimizations whose solutions cannot be easily characterized at the current level of generality when $X$ includes continuously distributed covariates and $\Pi$ may be arbitrarily restricted, which are both primary cases of interest of this paper. 
 To make progress, we impose the following restrictions on the ambiguity sets for the expected potential outcomes and CATE.
 \begin{assumption}[Rectangular identified set for $(y_0,y_1)$]\label{rectangularY}
The identified set for $(y_0,y_1)$ is rectangular, that is, $\mathcal{Y}$ is of the form
	\begin{equation*}
		\mathcal{Y} = \{ \left(y_0(\cdot),y_1(\cdot)\right) : \left(y_0(x),y_1(x)\right) \in \mathcal{Y}(x)\},
	\end{equation*}
	where $\mathcal{Y}(x)$ is a compact subset of $\mathbb{R}^2$.
\end{assumption}
\begin{assumption}[Rectangular identified set for $\tau$]\label{rectangularTau} The identified set for $\tau$ is rectangular, that is, $\mathcal{T}$ is of the form
	\begin{align*}
		\mathcal{T} = \{ \tau(\cdot) : \tau(x) \in [\underline{\tau}(x), \overline{\tau}(x) ]\},
	\end{align*}
	where $|\overline{\tau}(x)|<\infty$, $|\underline{\tau}(x)|<\infty$ for all $x\in\mathcal{X}$.
\end{assumption} Assumptions \ref{rectangularY} and \ref{rectangularTau} impose separation of the identified sets for the expected potential outcomes and CATE across the support of the covariates $\mathcal{X}$.\footnote{Notice that Assumption \ref{rectangularY} implies Assumption \ref{rectangularTau}, but not viceversa.} They are typically satisfied by identification schemes that do not impose shape restrictions on counterfactual outcomes with respect to the covariates $X_i$. These assumptions are widely adopted in the partial identification literature, and we refer the reader to Appendix B in \citet{kasy2016} for an extensive review of identification schemes that result in rectangular identified sets. Below we present three examples of identification schemes for the CATE that satisfy this assumption.

\begin{example}[Manski bounds]\label{manskiBounds}
	Suppose there exists a binary instrument $Z_i \in \{0,1\}$ that satisfies the well know exogeneity and exclusion restrictions $Y_i(0),Y_i(1),D_i(0),D_i(1) \perp Z_i | X_i$, where $Y_i(d)$ and $D_i(z)$ denote the counterfactual outcome and treatment functions, respectively. If the instrument $Z_i$ also satisfies the overlap condition
\begin{align*}
\eta \leq \mathbb{P}(Z_i =1 | X_i) \leq 1-\eta , \quad \eta>0,
\end{align*}	
	and the monotonicity condition (also known as no-defiers condition):
	\begin{align*}
		\mathbb{P} \big( D_i(1) \leq D_i(0)   | X_i\big) =1\quad {\rm or} \quad \mathbb{P} \big( D_i(1) \geq D_i(0)  | X_i\big) =1,
	\end{align*} 
	then seminal work by \citet{imbensangrist1994} shows point-identification of the conditional local average treatment effect (LATE): \begin{align*}\mathbb{E}[Y_i(1) - Y_i(0) \,| \, D_i(1) \neq D_i(0), \,X_i=x].\end{align*}
	Let us now assume that $Y\in [Y_L,Y_U]$, i.e.\@\ the outcome is bounded, and define 
	\begin{align*}
		&h(z,x) = \mathbb{E}[Y_i | Z_i=z, X_i=x],\\
		&m(d,z,x) = \mathbb{E}[Y_i |D_i=d, Z_i=z, X_i=x],\\
		&p(z,x)=\mathbb{P}(D_i=1 | Z_i=z, X_i=x),\\
		&z(x)= \mathbb{P}(Z_i=1 | X_i=x).
	\end{align*}
The identified sets for the expected potential outcomes $y_0(x)$ and $y_1(x)$ are contained within the bounds
\begin{align*}
	\overline{y}_0(x)& = \min_{z\in\{0,1\}}\big\{ m(0,z,x)\cdot (1-p(z,x)) +Y_U \cdot p(z,x)\big\}, \\
	\underline{y}_0(x)& = \max_{z\in\{0,1\}}\big\{ m(0,z,x)\cdot (1-p(z,x)) +Y_L \cdot p(z,x)\big\},	
\end{align*}
and 
\begin{align*}
	\overline{y}_1(x)& = \min_{z\in\{0,1\}}\big\{ m(1,z,x)\cdot p(z,x) +Y_U \cdot (1-p(z,x))\big\}, \\
	\underline{y}_1(x)& = \max_{z\in\{0,1\}}\big\{ m(1,z,x)\cdot p(z,x) +Y_L \cdot (1-p(z,x))\big\}.		
\end{align*}
The identified set for the CATE is then contained within the bounds
\begin{align*}
	\overline{\tau}(x)& = \overline{y}_1(x) - \underline{y}_0(x), \\
	\underline{\tau}(x) &=\underline{y}_1(x) -\overline{y}_0(x) .		
\end{align*}
	If no further functional form assumption on the distribution of potential outcomes is made, these bounds are sharp \citep{heckman2001instrumental} and the sharp identified sets for the average potential outcomes and CATE respectively satisfy Assumption \ref{rectangularY} and Assumption \ref{rectangularTau}.
\end{example}

\begin{example}[Balke-Pearl]\label{balkePearl}
Suppose that the same assumptions as in Example \ref{manskiBounds} hold, and additionally the monoticity assumption is strengthened to 
\begin{align*} \mathbb{P} \big( D_i(1) \geq D_i(0)  | X_i\big) =1,
\end{align*}
that is, the direction of the monotonicity is known and positive.
The bounds for the potential outcomes simplify to
\begin{align*}
\overline{y}_0(x)& = m(0,0,x)\cdot (1-p(0,x)) +Y_U \cdot p(0,x), \\
	\underline{y}_0(x)& =  m(0,0,x)\cdot (1-p(0,x)) +Y_L \cdot p(0,x),\\
	\overline{y}_1(x)& =  m(1,1,x)\cdot p(1,x) +Y_U \cdot (1-p(1,x), \\
	\underline{y}_1(x)& =  m(1,1,x)\cdot p(1,x) +Y_L \cdot (1-p(1,x)),
\end{align*}
and the CATE is contained within the bounds 
\begin{align*}
	\overline{\tau}(x)& = h(1,x) - h(0,x) + p(0,x)\cdot\big( m(1,0, x) - Y_L\big) + (1-p(1,x))\cdot\big(Y_U - m(0,1, x) \big), \\
	\underline{\tau}(x) &= h(1,x) - h(0,x) + p(0,x)\cdot\big( m(1,0, x) - Y_U\big) + (1-p(1,x))\cdot\big(Y_L - m(0,1, x) \big) .		
\end{align*}
	where $p(0,x)$ and $1-p(1,x)$ identify the proportions of always-takers and never-takers at $X_i=x$, respectively. If no further functional form assumption on the distribution of outcomes for non-compliant populations is made, these bounds are sharp  \citep{balkePearl2007} and the sharp identified sets for the average potential outcomes and CATE respectively satisfy Assumption \ref{rectangularY} and Assumption \ref{rectangularTau}.
\end{example}

\begin{example}[Manski-Pepper bounds]\label{ManskiPepper}
	Suppose that instead of full exogeneity, the instrumental variable $Z_i$ satisfies the weaker ``monotone IV" condition
	\begin{align}\label{MIV}
		\E[Y_i(d) | Z_i=0 , X_i ] \leq \E[Y_i(d) | Z_i=1 , X_i ] , \quad d=0,1.
	\end{align}
 \citet{manskiPepper} show that when the outcome is bounded one has
 \begin{align*}
 	\sum_{z=0,1}\mathbb{P}\left(Z_i=z|X_i \right)\cdot \max_{z_1\leq z} &\left\{ m(d,z_1,X_i)\cdot \mathbb{P}(D_i=d | Z_i=z_1, X_i) + Y_L\cdot \mathbb{P}(D_i=1-d|Z_i=z_1,X_i) \right\} \\
 	&\qquad \qquad \qquad \leq \,\E[Y_i(d) | X_i] \,\leq\\
 	\sum_{z=0,1}\mathbb{P}\left(Z_i=z|X_i \right)\cdot \min_{z_2\geq z}&\left\{ m(d,z_2,X_i)\cdot \mathbb{P}(D_i=d | Z_i=z_2, X_i) + Y_L\cdot \mathbb{P}(D_i=1-d|Z_i=z_2,X_i)\right\}.
 \end{align*} 
 
Upper (lower) bounds for the CATE are obtained by combining upper (lower) bounds for $\E[Y_i(1)|X_i=x]$ with the lower (upper) bound for $\E[Y_i(0) | X_i=x]$:
	\begin{align*}
		\overline{\tau}(x)& = z(x)\cdot \psi_{1,1}(x;Y_U) + (1-z(x))\cdot \min\left\{\psi_{0,1}(x;Y_U),\psi_{1,1}(x;Y_U)\right\} \\
		&-z(x)\cdot\max\left\{\psi_{0,0}(x;Y_L),\psi_{1,0}(x;Y_L)\right\}  -(1-z(x))\cdot \psi_{0,0}(x;Y_L), \\
		\underline{\tau}(x) &= z(x)\cdot \max\left\{\psi_{0,1}(x;Y_L),\psi_{1,1}(x;Y_L)\right\} + (1-z(x))\cdot \psi_{0,1}(x; Y_L)\\
		&-z(x)\cdot \psi_{1,1}(x;Y_U) -(1-z(x))\cdot \min\left\{\psi_{0,0}(x;Y_U),\psi_{1,0}(x;Y_U)\right\}, .		
	\end{align*}
where 
\begin{align*}
	\psi_{z,d}\left(x;Y_{(\cdot)}\right)=m(d,z,x)\cdot (d\cdot p(z,x) +(1-d)\cdot (1-p(z,x))  + Y_{(\cdot)} \cdot (d\cdot (1-p(z,x)) +(1-d)\cdot p(z,x)).
\end{align*}
Under no further assumption on the distribution of potential outcomes, these bounds are sharp \citep{manskiPepper} and satisfy Assumptions \ref{rectangularY} and \ref{rectangularTau}.
\end{example}

Having restricted the identified sets $\mathcal{Y}$ and $\mathcal{T}$ as in Assumptions \ref{rectangularY}-\ref{rectangularTau}, we are now able to provide a simpler characterization of the maximin welfare and maximin impact policies.
\begin{proposition}\label{prop:ReducedMaxmin}
	Define $\underline{y}_d(x)=\min_{y_d(x) \in \mathcal{Y}(x)} y_d(x)$ and $\overline{y}_d(x)=\max_{y_d(x) \in \mathcal{Y}(x)} y_d(x)$. Under Assumption \ref{rectangularY} the optimal maximin welfare policy is
	\begin{align}\label{ReducedMaxminWelfare}
		\pi^*_{\emph{\texttt{MMW}}} = \argmax_{\pi \in \Pi } \mathbb{E}_{P_X}\Bigg[ \left( 2\pi(X)-1\right) \cdot (\underline{y}_1(X) -\underline{y}_0(X)) \Bigg].
	\end{align}
Furthermore, under Assumption \ref{rectangularTau} the optimal maximin impact policy is
	\begin{align}\label{ReducedMaxminImpact}
		\pi^*_{\emph{\texttt{MMI}}} = \argmax_{\pi \in \Pi } \mathbb{E}_{P_X}\Bigg[ \left( 2\pi(X)-1\right) \cdot \underline{\tau}(X) \Bigg].
	\end{align}
\end{proposition}
Proposition \ref{prop:ReducedMaxmin} shows that the optimal maximin welfare and maximin impact policies maximize surrogate versions of the welfare substituting the unidentified CATE with the difference in the lower bounds of potential outcomes $\underline{y}_1(x)-\underline{y}_0(x)$ and the lower bound for CATE $\underline{\tau}(x)$ at every point in the covariate space, respectively. Notice that when $\mathcal{Y}(x)$ is rectangular with respect to the two potential outcomes, i.e.\@ $y_0(x)\in \mathcal{Y}_0(x), y_1(x)\in \mathcal{Y}_1(x)$ and $\mathcal{Y}(x)=\mathcal{Y}_0(x)\times\mathcal{Y}_1(x)$, we have $\underline{\tau}(x)=\underline{y}_1(x)-\overline{y}_0(x)$, thus highlighting the ``pessimistic" nature of the maximin impact criterion.
\begin{remark}
The maximin welfare and maximin impact optimal policies coincide when $y_0(\cdot)$ is point-identified. This case is relevant when $y_0(x)$ represents the (conditional) average outcome under the status-quo in the entire population and is typically point-identified from observational data.
\end{remark}
The simplification of these two maximin problems into single maximisation problems has important advantages for the study of the optimal policies and their estimation from the data. In fact, the sample analogues of optimizations \eqref{ReducedMaxminWelfare} and \eqref{ReducedMaxminImpact} are amenable to standard computation procedures for a variety of policy classes $\Pi$. Furthermore, their solution can be studied using tools for empirical risk minimisation problems, as discussed in Section \ref{sec:Estimation}.

Despite the involvement of an additional maximization problem compared to maximin welfare and maximin impact, Assumption \ref{rectangularTau} allows to provide a simpler characterization also for the minimax regret optimal policy.
\begin{proposition}\label{prop:minimaxRegret}
	Under Assumption \ref{rectangularTau} the optimal minimax welfare regret policy is
	\begin{align}\label{PopulationPolicy}
		\pi^*_{\emph{\texttt{MMR}}} = \argmax_{\pi \in \Pi } \mathbb{E}_{P_X}\Bigg[ \left( 2\pi(X)-1\right) \cdot \widetilde \tau(X) \Bigg]
	\end{align}
where 
\begin{align}\label{scoresMMR}
	 \widetilde \tau(x)=\overline{\tau}(x)\cdot\mathbbm{1}\big\{\overline{\tau}(x) \geq0 \big\} + \underline{\tau}(x)\cdot \mathbbm{1}\big\{\underline{\tau}(x) \leq0 \big\}
\end{align}
\end{proposition}
This simpler characterization of the minimax regret problem as a single maximization sheds light on the properties of its associated optimal policy. In particular, we see that the objective function symmetrically treats individuals whose expected treatment effect sign is identified by assigning as surrogate for the CATE their outer bound, i.e.\@\ the CATE upper (lower) bound for individuals with identified positive (negative) sign for CATE. Individuals for which the sign of the treatment effect is ambiguous are assigned an intermediate point within their respective CATE bounds. The location of this intermediate point depends on the extent to which the identified set lies in the positive or negative region. Intuitively, the criterion prioritizes correct treatment allocation to individuals who unambiguously benefit from (or are harmed by) the treatment and down-weights the importance of individuals for which the sign of the treatment response is ambiguous within the treatment allocation problem. As an extreme case, individuals with CATE bounds exactly symmetric around 0 (i.e.\@ $\overline{\tau}(x)=-\underline{\tau}(x)$) are given no consideration in the solution of the treatment allocation problem. This intuition can be further supported by noticing that the original welfare maximization under point-identification in (\ref{generalOptimalPolicy}) can be re-casted as the weighted classification problem
\begin{align*}
	\pi^*= \argmin_{\pi \in \Pi}\E_{P_X}\bigg[\mathbbm{1}\big\{ (2\pi(X) -1) \neq \text{sign}(\tau(X))\big\}\cdot|\tau(X)|\bigg],
\end{align*}
of which the minimax welfare regret optimal policy in \eqref{scoresMMR} turns out to solve the minimax version under Assumption \ref{rectangularTau}:
\begin{align*}
\pi^*_{\texttt{MMR}} =  \argmin_{\pi \in \Pi}\max_{\tau \in \mathcal{T}}\E_{P_X}\bigg[\mathbbm{1}\big\{ (2\pi(X) -1) \neq \text{sign}(\tau(X))\big\}\cdot|\tau(X)|\bigg].
\end{align*}
It is from this minimax classification risk perspective that \citet{Pu_2021} obtain and study the minimax regret policy, which they call the ``IV-optimal policy". 

An alternative version of minimax regret optimality which has been used in the context of treatment choice is minimax regret with respect to a baseline policy. \citet{kallus2018} assume the existence of a fixed policy $\pi_{\texttt{B}}$ from which the DM does not want to unnecessarily deviate. They define the optimal policy as minimizing regret with respect to this baseline policy:
\begin{align*}
		\pi^*_{\texttt{MMRB}}&=\argmin_{\pi \in \Pi } \max_{\tau \in \mathcal{T} }\left( I_\tau(\pi_{\texttt{B}}) - I_\tau(\pi)\right)\\
		&=\argmax_{\pi \in \Pi } \mathbb{E}_{P_X}\Bigg[ \left( 2\pi(X)-1\right) \cdot \left(\overline{\tau}(X)\cdot\mathbbm{1}\left\{\pi_{\texttt{B}}(X)\geq0\right\}+\underline{\tau}(X)\cdot\mathbbm{1}\left\{\pi_{\texttt{B}}(X)<0\right\} \right) \Bigg],
\end{align*}
where the second equality uses Assumption \ref{rectangularTau}. While potentially appealing in certain settings, e.g.\@ when $\pi_{\texttt{B}}$ represents the existing standard of care in a medical setting, this optimality criterion suffers the potential drawback of requiring the DM to specify (and motivate) the baseline policy for it to be operational. Adopting the never-treat baseline policy, i.e.\@ $\pi_{\texttt{B}}(x)=0, \, \forall x\in\mathcal{X}$, could be seen as an appealing ``agnostic" choice, which however makes this criterion default to maximin impact and thus inherit its potentially undesirable properties.

The last notion of ambiguity-robust optimal policy that we present in this section is based on the Hurwicz criterion \citep{hurwicz}, arguably one of the most widely used in decision-making under ambiguity. In the context of the treatment assignment problem at hand, the Hurwicz criterion leads to the ambiguity-robust policy
	\begin{align*}
		&\pi^*_{\texttt{HurW},\delta_0,\delta_1} =\\& \argmax_{\pi \in \Pi } \mathbb{E}_{P_X}\Bigg[ \left( 2\pi(X)-1\right) \cdot \big(\{\delta_1\cdot\overline{y}_1(X) +(1-\delta_1)\cdot\underline{y}_1(X)\} -\{\delta_0\cdot\overline{y}_0(X) +(1-\delta_0)\cdot\underline{y}_0(X)\}\big) \Bigg],
	\end{align*}
where $\delta_1\in[0,1]$ and $\delta_0\in[0,1]$ are user-defined weights reflecting the degree of optimism with respect to the outcomes under treatment and non-treatment, respectively. It is easy to see that the maximin welfare criterion in \eqref{maximinWelfare} corresponds to the choice $\delta_1=0,\delta_0=0$. An analogous notion of optimality focused on impact rather than welfare, leads to the optimal policy
\begin{align*}
		\pi^*_{\texttt{HurI},\delta} =\argmax_{\pi \in \Pi } \mathbb{E}_{P_X}\Bigg[ \left( 2\pi(X)-1\right) \cdot \big(\delta\cdot\overline{\tau}(X) +(1-\delta)\cdot\underline{\tau}(X)\big) \Bigg],
	\end{align*}
where $\delta\in[0,1]$ controls the degree of optimism with respect to the effect of treatment, with the maximin impact optimal policy corresponding to the choice $\delta=0$. Under Assumption \ref{rectangularY} and $\mathcal{Y}(x)=\mathcal{Y}_0(x)\times\mathcal{Y}_1(x)$, the Hurwicz Impact criterion is nested into the Hurwicz Welfare for the choice of parameters $\delta=\delta_0=1-\delta_1$; unlike Hurwicz Welfare, however, the Hurwicz Impact criterion is still well-defined under the weaker Assumption \ref{rectangularTau}.  Interestingly, when $\delta=1/2$ and $\Pi$ is well-specified, in the sense that it contains the first-best assignment $\mathbbm{1}\{\overline{\tau}(x) + \underline{\tau}(x)\geq 0 \}$, we have
\begin{align*}
\pi^*_{\texttt{MMR}}(x)=\mathbbm{1}\{\widetilde \tau(x)\geq 0\}=\mathbbm{1}\{\overline{\tau}(x) + \underline{\tau}(x)\geq 0\}=\pi^*_{\texttt{HurI},\frac{1}{2}}(x).
\end{align*}
Therefore the minimax regret and Hurwicz impact optimal policies coincide under correct specification of $\Pi$, as they assign treatment based on the middle point between the upper and lower CATE bounds.
 When $\Pi$ is not well-specified, however, minimax regret optimality is not nested into any of the Hurwicz-type criteria just presented, thus highlighting the radically different attitude towards ambiguity implied by minimax regret compared to maximin welfare/impact. In particular, minimax regret is the only criterion of those presented (along with Hurwicz impact under $\delta=1/2$) that treats symmetrically individuals with identified sets symmetric around 0, in the sense that $\widetilde\tau_1(x)= -\widetilde\tau_2(x)$ whenever $\overline{\tau}_1(x)=-{\underline{\tau}}_2(x)$ and $\underline{\tau}_1(x)=-{\overline{\tau}}_2(x)$. For this reason, minimax regret does not reflect an optimistic/pessimistic attitude towards ambiguity but rather an ``opportunistic" one, in light of its prioritization of correct treatment assignment to individuals whose CATE sign is unambiguously identified.

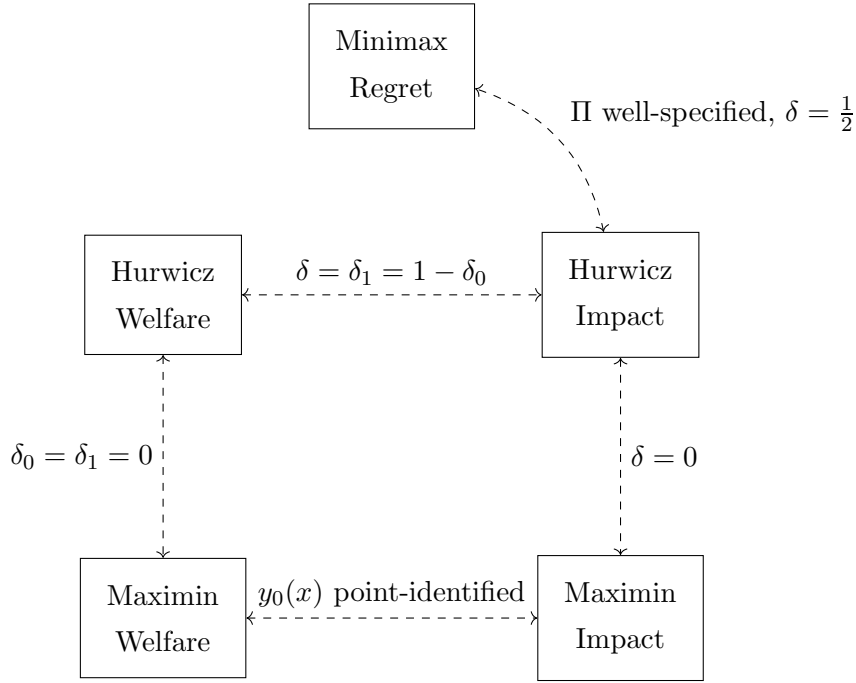
\begin{figure}[H]
	\centering
	\begin{tikzpicture} [node distance = 4.3cm, on grid, auto, ->,-=stealth']
 
 
\node (q0) [draw,inner sep=10pt, align=center]  {Minimax\\ Regret};
\node (q1) [draw,inner sep=10pt,below left = of q0, align=center] {Hurwicz\\ Welfare};
\node (q2) [draw,inner sep=10pt, below right = of q0,align=center] {Hurwicz\\ Impact};

\node (q3) [draw,inner sep=10pt, below  = of q1,align=center] {Maximin\\ Welfare};
\node (q4) [draw,inner sep=10pt, below = of q2,align=center] {Maximin \\ Impact};

\path 
   (q1) edge [<->, dashed]  node[-, dashed,left]   {$\delta_0=\delta_1=0$} (q3)
    (q2) edge [<->, dashed,bend right] node[above right] {$\Pi$ well-specified, $\delta=\frac{1}{2}$}  (q0);
    
\path 
    (q2) edge [<->, dashed] node[right] {$\delta=0$} (q4);


 \path (q3) edge [<->, dashed] node [above] {$y_0(x)$ point-identified} (q4);
 
 \path (q1) edge [<->, dashed] node [above] {$\delta=\delta_1=1-\delta_0$} 
 (q2);
 

\end{tikzpicture}
	\caption{Relationship between ambiguity-robust optimal policies}\label{fig:ConceptualMap}
\end{figure}

\subsection{A common framework}
While accommodating a wide range of attitudes towards ambiguity, the notions of optimality presented in Section \ref{subsec:RegretConvergenceRates} share a common structure. In fact, by virtue of Assumptions \ref{rectangularY} and \ref{rectangularTau}, the corresponding optimal policies can all be written as
\begin{align}\label{formOfPolicies}
	\pi^*(P)=\argmax_{\pi \in \Pi } Q(P;\pi),\qquad  Q(P;\pi):=\mathbb{E}_{P_X}\Big[ \left( 2\pi(X)-1\right) \cdot \Gamma(P;X) \Big],
\end{align} 
for a specific score function\footnote{The term `score function' is borrowed from \citet{atheywager2021}.} $\Gamma(P;\,\cdot\,)$, where we have highlighted the dependence of the score on the distribution $P$. The specific dependence on $P$ is determined by the optimality criterion (as summarized in Table \ref{tableScores}) as well as the identification assumptions (e.g.\@ Balke-Pearl, Manski-Pepper etc.\@). This common structure also nests the point identified setting as the special case $\Gamma(P;X)=\tau(X)$ and thus suggests that existing estimation procedures for this special case can be extended to the partially identified setting. 

However, one peculiar feature of the partially identified setting is the restricted degree of smoothness enjoyed by the objective function, in particular the differentiability of the scores with respect to $P$. Under point-identification of the CATE via standard unconfoundedness assumptions, one has $\Gamma(P;x)=\mathbb{E}[Y| D=1,X=x]-\mathbb{E}[Y|D=1,X=x]$ and the full differentiability of the score with respect to the expectation $\E[Y|D,X]$ is immediately apparent. However, for the minimax regret criterion we notice that the score is  \textit{directionally differentiable}\footnote{ Let $P\in\mathcal{P}$ be a probability distribution on which the function $f:\mathcal{P}\to \mathbb{R}$ depends. We say that $f$ is directionally differentiable at $P_0$ if the limit \begin{align*}
\lim_{t \downarrow 0} \frac{f(P_0 +t(h-P_0))-f(P_0)}{t}=\dot{f}_{P_0}[h]
\end{align*} exists for every $h\in \mathcal{P}$, in which case $\dot{f}_{P_0}[\cdot]$ denotes the directional derivative of $f$ at $P_0$. If it exists, the directional derivative $\dot{f}_{P_0}[\cdot]$ is positively homogeneous of degree one but not
necessarily linear. If $\dot{f}_{P_0}[\cdot]$ is linear then f is fully differentiable at $P_0$.} with respect to $P$ at $\overline{\tau}(x)=0$ or $\underline{\tau}(x)=0$. Even when $\Gamma(P;x)$ depends smoothly on expected outcomes/CATE bounds, lack of full differentiability of the scores can arise through a lack of differentiability of the expected outcomes and CATE bounds themselves. In fact, many popular identification assumptions, including the Manski and Manski-Pepper bounds from Examples \ref{manskiBounds} and \ref{ManskiPepper}, deliver bounds that are only directionally differentiable with respect to identified parameters due to the presence of $\min/\max$ operators \citep[see][and examples therein]{chernozhukovIntersectionBounds}. Whether a consequence of the optimality criterion or the identification assumptions, lack of full differentiability of the scores is a unique and pervasive feature of the treatment assignment problem under partial identification, one that has not been explicitly acknowledged in the most recent contributions in this area.\footnote{The only exception is \cite{christensenMoonSchor2020}, who deal with estimation of optimal treatment decisions in the absence of individualization.} A major contribution of this paper is to account for the role played by the lack of full differentiability when we establish procedures for estimating ambiguity-robust optimal policies in Section \ref{sec:Estimation}. 

\begin{table}[H]
\caption{Optimality criteria and associated scores}\label{tableScores}
	\centering
	\renewcommand{\arraystretch}{1.1}
	\begin{threeparttable}
	\begin{tabular}[t]{lccc}
		
		Optimality criterion & $\Gamma(P;x)$\\
\midrule\midrule
		Maximin Welfare	&$\underline{y}_1(x) - \underline{y}_0(x)$\\
		Maximin Impact	& $\underline{\tau}(x)$\\
		Minimax Regret (oracle) & $\overline{\tau}(x)\cdot\mathbbm{1}\big\{\overline{\tau}(x) \geq0 \big\} + \underline{\tau}(x)\cdot \mathbbm{1}\big\{\underline{\tau}(x) \leq0 \big\}$\\
		Minimax Regret (baseline) & $\overline{\tau}(x)\cdot \mathbbm{1}\big\{\pi_{\texttt{B}}(x)=1\big\} +\underline{\tau}(x)\cdot\big\{\pi_{\texttt{B}}(x)=0\big\} $\\
		Hurwicz (welfare)	& \makecell{$\big(\{\delta_1\cdot\overline{y}_1(x) +(1-\delta_1)\cdot\underline{y}_1(x)\} -\{\delta_0\cdot\overline{y}_0(x) +(1-\delta_0)\cdot\underline{y}_0(x)\}\big)$\\ $\delta_1,\delta_0\in[0,1]$}\\
		Hurwicz (impact)	& $\delta\cdot\overline{\tau}(x) + (1-\delta)\cdot \underline{\tau}(x), \, \delta\in[0,1]$\\
\bottomrule
	\end{tabular}
\end{threeparttable}
\end{table}%
\section{Estimation}\label{sec:Estimation}
In this section we present the statistical framework underlying the problem of estimation of optimal treatment rules under partial identification. We will discuss heuristics underlying several features of the estimation problem, and then present our proposed estimation procedure.

We work in a learning setting where the estimand $\pi^*(P)$ is as in (\ref{formOfPolicies}), and we observe an i.i.d. sample $(W_i)_{i=1,\dots,n}$ of size $n$ from the unknown distribution $P$ of the observed random variables $W\in \mathcal{W}$, $X\subset W$. To retain generality of the framework, we do not specify the exact dependence of the functional $\Gamma(P;x)$ on $P$, which will depend on the choice of optimality criterion for the resolution of ambiguity (maximin welfare, minimax regret etc.) and identification assumptions determining the identification sets  $\mathcal{Y}(P),\mathcal{T}(P)$. However, we will assume that the scores depend on $P$ only through a vector of nuisance functions $g:\mathcal{V}\to \mathbb{R}^J$ specified by the moment equations
\begin{align}\label{momentEquations}
\E[U - g(V) \mid V ] =0,
\end{align} 
where $U$ and $V$ are random vectors with $U\subseteq W$ and $X\subseteq V\subset W$. Furthermore, we will stipulate that the dependence of $\Gamma(g;x)$ on the nuisance functions $g$ from the possibly infinite-dimensional space $\mathcal{G}$ can be reduced as 
\begin{align*}
\Gamma(g;x)=\Gamma(\theta(x),x),
\end{align*}
where, for a fixed $x$, the parameter $\theta(x)\in \Theta_x \subseteq\mathbb{R}^M$ is a finite-dimensional vector of conditional moments of $U$ deduced from $g$. This latter restriction rules out scores $\Gamma(g;X)$ that at a single point in the covariate space depend on exhaustive evaluations of the nuisance functions $g$ over continuous supports. This is the case, for example, in versions of the CATE bounds from Examples \ref{manskiBounds}-\ref{ManskiPepper} featuring instruments with continuous support $\mathcal{Z}$. In those settings, the CATE bounds depend on objects such as $\sup_{z\in \mathcal{Z}}\E[Y \mid Z=z, \,X=x ]$, and are therefore not covered by the results of this paper.
Finally, we will assume that that $\Gamma(\theta(x);x)$ can be expressed as 
\begin{equation}\label{functionalFormScores}
\begin{aligned}
\Gamma(\theta(x);x) =\varphi_0(\theta(x);x) + \sum_{\ell=1}^L a_{\ell}\cdot\varphi_\ell(\theta(x);x)\cdot \mathbbm{1}\left\{\varphi_\ell(\theta(x);x)\geq 0 \right\}, \quad a_\ell\in\{-1,1\},
\end{aligned}
\end{equation}
where the functions $\varphi_\ell(\theta(x);x): \Theta_x\times \mathcal{X} \to \mathbb{R}$ are fully differentiable with respect to $\theta(x)$ for all $x\in\mathcal{X}$. While seemingly ad-hoc, this restriction is sufficiently general to 
accommodate a wide range of popular partial identification assumptions for the CATE as well as optimality criteria for the resolution of ambiguity. In particular, formulation (\ref{functionalFormScores}) accommodates linear combinations of $\min$/$\max$ operators, which typically feature in many identification bounds for the CATE with discrete instruments. In fact, our framework can be shown to be applicable to any combination of the optimality criteria discussed in Section \ref{sec:policyDerivation} and the identification schemes contained in the recent survey paper by Swanson et al. (2018).\footnote{Albeit not directly accommodated by formulation \eqref{functionalFormScores}, our framework and theoretical results also apply to scores that feature a finite number of nested linear combinations of $\min/\max$ operators. We discuss this extension in Appendix \ref{appendixExtensionMinMax}.}
\begin{example}[continues=balkePearl]
Under the identification assumptions of the Balke-Pearl bounds and resolution of ambiguity via Minimax Regret, we have \begin{align*}g&=(h, m, p),\\ \theta(x)&=(h(1,x), h(0,x), m(1,0,x), m(0,1,x), p(1,x), p(0,x)),\end{align*}
and
\begin{align*}
\Gamma(g;x)= \varphi_1(\theta(x);x)\cdot \mathbbm{1}\left\{\varphi_1(\theta(x);x)\geq 0 \right\} - \varphi_2(\theta(x);x)\cdot \mathbbm{1}\left\{\varphi_2(\theta(x);x)\geq 0 \right\},
\end{align*}
where
$\varphi_1(\theta(x);x)=\overline{\tau}(\theta(x);x)$, $\varphi_2(\theta(x);x)=-\underline{\tau}(\theta(x);x)$ are differentiable with respect to $\theta(x)$.
\end{example}

In this framework, a natural approach for estimation is via the so-called ``empirical risk minimisation" (ERM) principle \citep{vapnik1998}, in which the estimate for the optimal policy is obtained as the maximiser of a sample analogue of the population objective $Q$:
\begin{align}\label{estimationPolicy}
	\widehat{\pi}_{n} = \argmax \Big\{\widehat Q_n(\pi) \, : \, \pi \in \Pi \Big\}, \qquad \widehat Q_n(\pi) = \frac{1}{n}\sum_{i=1}^n(2\pi(X_i)-1)\cdot\widehat \Gamma_i
\end{align}
where $\widehat{\Gamma}_i$ is some suitable estimate for $\Gamma(g;X_i)$. The ERM approach is a cornerstone of statistical learning theory and is at the foundation of many traditional and modern estimation methods in statistics, econometrics and machine learning. The ERM principle has also guided much of the recent literature on individualized treatment rules, where different variations have been applied under the names of ``outcome-weighted learning" \citep{kosorokOWL2012} and ``empirical welfare maximization" \citep{kitagawatetenov2018}. A major challenge in the implementation of \eqref{estimationPolicy} comes from the presence of the nuisance functions $g$, which are typically unknown and thus need to be estimated. Assuming that we have access to appropriate algorithms/nonparametric procedures for estimation the nuisance functions, one simple approach would be to use the sample
$(W_i)_{i=1,\dots,n}$ to obtain the estimates $\widehat g$ and then form plug-in estimates for the score as $\widehat \Gamma_i=\Gamma(\widehat g; X_i)$. While seemingly natural, this ``naive plug-in" approach has undesirable properties. In particular, policies estimated via the naive plug-in approach can typically only be shown to converge at sub-optimal rates to their population counterparts, unless very restrictive assumptions are imposed on first-stage estimators for the nuisance functions $g$ \citep[see, e.g.,][]{fosterSyrgkanis}. 

One crucial reason 
underlying the undesirable statistical properties of the naive plug-in approach is that the resulting objective function estimate $\widehat Q_n$ is overtly sensitive to error in estimating the nuisance functions $g$. In order to gain intuition, it is useful to consider the following expansion of the population objective function $Q(g;\pi)=\E_{P_X}\left[(2\pi(X)-1)\cdot\Gamma(g;X)\right]$,
\begin{align}\label{eq:taylorExpansion}
Q(\widetilde g;\pi) - Q(g;\pi) =
\frac{\partial  Q(g+t(\widetilde g - g);\pi)}{\partial t}\biggr\rvert_{t = 0} +\Delta(\widetilde g, g; \pi)
+ O\left(\|\widetilde g-g\|_{L_2(P)}^2\right)
\end{align}
where
\begin{align*}&\Delta(\widetilde g, g; \pi)=\\&\mathbb{E}_{P_X}\left[(2\pi(X)-1)\cdot\left(\sum_{\ell=1}^L a_\ell \cdot \varphi_\ell(g;X)\cdot \left(\mathbbm{1}\left\{\varphi_\ell(\widetilde g;X)\geq 0 \right\}-\mathbbm{1}\left\{\varphi_\ell(g;X)\geq 0 \right\}\right)\right)\right]. \end{align*}
Von Mises expansions like the above are at the heart of the theory of orthogonal machine learning \citep{chernozhukov2018}. In our setting, it allows to describe the impact of a small deviation from $g$ in the direction $\widetilde g -g$ as consisting of three terms. The first term is the so-called ``pathwise derivative" of $Q(g;\pi)$ and typically scales with $\|\widetilde g - g\|_{L_1(P)}$. The second term $\Delta(\widetilde g, g; \pi)$ is due the presence of the type of non-differentiabilities arising under partial identification, and is unique to the framework of this paper. 
This term accounts for the bias that arises from misclassifying whether the component functions $\varphi_\ell$ are above or below 0, as we move away from $g$ in the direction $\widetilde g - g$.
The third term is a second-order remainder scaling with the mean-square distance between $\widetilde g$ and $g$. A central feature of our proposed estimation procedure is the construction of a new objective function, called Neyman-orthogonal, with reduced sensitivity to local perturbations away from $g$. For this purpose, we will assume that there exists functionals $\alpha_\ell(\{g,f\}; V)$ such that for every $\widetilde g \in \mathcal{G}$
\begin{align*}
\varphi_\ell(\widetilde g;x) = \E[\langle\alpha_\ell(\{\widetilde g, f\};V), \widetilde g(V) \rangle \mid X=x], \qquad \ell=1,\dots,L,
\end{align*}
where $f\in \mathcal{F}$ is a vector of additional nuisance functions defined analogously to $g$
.\footnote{We will also assume that for the $j$-th entry of the Riesz-representer we have $\alpha^{(j)}_\ell(\{(\widetilde g_{-j},\widetilde g_{j}),\widetilde f\}, x)=\alpha^{(j)}_\ell(\{\widetilde g_{-j},\widetilde f\}, x)$, where $\widetilde g_{-j}$ denotes the exclusion of the $j$-th entry $\widetilde g_{j}$ from the vector of nuisance functions $\widetilde g$. This restriction is sufficiently general to accommodate  component functions $\varphi_\ell(\theta(x);x)$ that feature linear combinations of products of the parameters $\theta(x)$, thus encompassing all the discussed identification schemes, including Examples \ref{manskiBounds}-\ref{ManskiPepper}.} We then construct Neyman-orthogonal formulations for the component functions as
\begin{align*}
\varphi_\ell^{\texttt{NO}}(\{g,f\};w) = \varphi_\ell(\theta;x) + \phi_\ell(\{g,f\};w), \quad \phi_\ell(\{g,f\}; w)=\langle\alpha_\ell(\{g,f\},v), u - g(v) \rangle.
\end{align*}
The functionals $\alpha_\ell$ are the Riesz-representers of $\varphi_\ell$, while the functionals $\phi_\ell$ are their so-called influence function adjustments. We refer the reader to \citet{ichimura2015} for their properties and general methods for their calculation, while we provide below their specific form for the Balke-Pearl CATE bounds of Example \ref{balkePearl}.\footnote{See also \citet{kennedy2022} for a user-friendly discussion of methods for computation influence function adjustments.}
\begin{example}[continues=balkePearl]
	Following \citet{ichimura2015}, we compute the influence function adjustment $\phi_U(\{g,f\},W_i)$ for the CATE upper bound by taking the Gateaux derivative of $\overline{\tau}(g;X)$, which yields
	 \begin{align*}
			\phi_U(\{g,f\};W_i) =&\underbrace{\left[\frac{Z_i}{z(X_i)} - \frac{1-Z_i}{1-z(X_i)}\right]}_{\alpha_U^{(1)}(\{g,f\},V_i)}\cdot\big(Y_i - h(Z_i,X_i)\big)\\
		&+\underbrace{\left[\frac{D_i(1-Z_i)}{1-z(X_i)}+\frac{(1-D_i)Z_i}{z(X_i)}\right]}_{\alpha_U^{(2)}(\{g,f\},V_i)}\cdot \big( Y_i -m(D_i,Z_i,X_i)\big)		\\
		&+\underbrace{\left[(m(1,0, X_i)-Y_L)\cdot \frac{1-Z_i}{1-z(X_i)} - (Y_U-m(0,1, X_i))\cdot \frac{Z_i}{z(X_i)}\right]}_{\alpha_U^{(3)}(\{g,f\},V_i)}\cdot \big(D_i - p(Z_i,X_i)\big),
	\end{align*}
	where the associated Riesz-representer is $\alpha_U(\{g,f\},V_i)=(\alpha_U^{(1)},\alpha_U^{(2)},\alpha_U^{(3)})'$ with $f=z(x)$ and $V_i=(D_i,Z_i,X_i)'$. The influence function and Riesz-representer for the CATE lower bound are obtained by interchanging $Y_U$ and $Y_L$ in the expressions above.
\end{example}
Finally we construct Neyman-orthogonal formulations for the scores as
\begin{align*}
\Gamma^{\texttt{NO}}(\{g,f\};w) = \varphi_0^{\texttt{NO}}(\{g,f\} ;w) + \sum_{\ell=1}^L a_{\ell}\cdot\varphi^{\texttt{NO}}_\ell(\{g,f\};w)\cdot \mathbbm{1}\left\{\varphi_\ell(g;x)\geq 0 \right\},
\end{align*}
which are then used to form the Neyman-orthogonal objective function
\begin{align*}
Q^{\texttt{NO}}(\{\cdot,\cdot\};\pi) = \mathbb{E}_{P_W}\Big[ \left( 2\pi(X)-1\right) \cdot \Gamma^{\texttt{NO}}(\{\cdot,\cdot\};W) \Big].
\end{align*}
Our construction of Neyman-orthogonal scores features the addition of the influence function adjustments $\phi_\ell$ to the component functions $\varphi_\ell$ outside the indicators, but crucially not inside. Heuristically, the influence function adjustments serve the purpose of reducing the bias induced by the evaluation of the component functions $\varphi_\ell(\cdot;x)$ away from $g$. Since the indicators vary discontinuously with $g$ it is not possible to linearly approximate the dependence of the indicators on the nuisance functions at the point of discontinuity $\varphi_\ell(g;x)=0$. As a result, it is not possible to reduce the bias induced by the presence of the indicators (represented by the term $\Delta(\widetilde g,g,\pi)$ in \eqref{eq:taylorExpansion}) by means of influence function adjustments, whose de-biasing properties implicitly rely on the validity of such linear approximation.\footnote{On the contrary, naively adding the influence function adjustments inside the indicators would lead to a bias increase, rather than a reduction.} Notice that $Q^{\texttt{NO}}(\{g,f\};\pi) = Q(g;\pi)$ by the mean-zero property of the influence function adjustments, so that orthogonalization of the objective does not change the notion of optimal policy $\pi^*(P)$. Nonetheless, for the orthogonalized objective we have that
\begin{align}\label{eq:taylorExpansion_orthogonal}
Q^{\texttt{NO}}(\{\widetilde g,\widetilde f\};\pi) - Q^{\texttt{NO}}(\{g,f\};\pi) =
\Delta(\widetilde g, g; \pi)
+ O\left(\| \widetilde g -g\|_{L_2(P)}^2 + \| \widetilde f -f\|_{L_2(P)}^2\right).
\end{align}
Comparing the above with (\ref{eq:taylorExpansion}), we see that the von Mises expansion for the orthogonalized objective does not feature the pathwise derivative term, implying that $Q^{\texttt{NO}}(\,\cdot\,;\pi)$ is less sensitive to deviations away from $g$ compared to the original objective $Q( \, \cdot \,;\pi)$. As shown in Section \ref{sec:Theory}, this property will generally translate in improved statistical guarantees for the estimated policy when the nuisance functions have to be learned from the data. It should however be noticed that the term $\Delta(\widetilde g,g;\pi)$ still appears in the relevant expansion after orthogonalization. The contribution of this term is quantified in Section \ref{sec:Theory}, where it is shown to be of first-order importance for the statistical properties of the estimation procedures.

The second key component of our approach is the use of sample-splitting, which is a commonly employed method in semiparametric inference \citep{chernozhukov2018} and statistical learning \citep{fosterSyrgkanis}. The main purpose of sample-splitting is to reduce the risk of overfitting that generally arises from using the same data to estimate the nuisance functions as well as the optimal policy, as in the naive plug-in approach. Similarly to \citet{atheywager2021}, we employ a particular form of sample-splitting known as K-fold cross-fitting (described below). This procedure ensures that in the estimate for $\Gamma^{\texttt{NO}}(\{g,f\}; W_i)$, the estimates for the nuisance functions $\{g,f\}$ are independent from the data-point $W_i$ for that same unit. This independence property is crucial for the theoretical guarantees of our proposed method.

Our proposed estimation procedure is therefore as follows. We first randomly split the data into $K$ evenly-sized folds and for each fold $k=1,\dots, K$ we obtain estimates $\{\widehat g^{(-k)},\widehat{f}^{(-k)}\}$ using data from the remaining $K-1$ folds. These estimates are then used to form cross-fitted Neyman-orthogonal estimates for the scores
\begin{align}\label{OrthogonalScores}
\widehat{\Gamma}_i = \widehat{\Gamma}^{\texttt{NO}}\left(\{\hat{g}^{-k(i)},\hat{f}^{-k(i)}\};W_i\right), \qquad i=1,\dots,n,
\end{align}
where $k(i)\in \{1,\dots, K\}$ denotes the fold containing the $i$-th observation. Finally, the estimated optimal policy rule $\widehat{\pi}_n$ is obtained via the optimization problem (\ref{estimationPolicy}). 

\section{Statistical guarantees for the estimated policy}\label{sec:Theory}
Let $\widehat{\pi}_{n}$ be the estimated treatment policy defined in (\ref{estimationPolicy}), with estimated scores as in (\ref{OrthogonalScores}). Following \citet{manski2004}, we assess the performance of the estimated policy in terms of (statistical) regret with respect to population optimal policy. 
Let the population ambiguity-robust optimal policy be $\pi^*_{n}(P)\in \argmax_{\pi \in \Pi_n }Q(P;\pi)$, where we have included the $n$-subscript to the policy class $\Pi_n$ to allow this to depend on the sample size for generality. The statistical regret of an estimated policy $\widehat{\pi}_n$ is defined as
\begin{align}\label{regret}
	R_{n}(P;\widehat{\pi}_n) = \mathbb{E}_{P_n}\big[Q(P;{\pi}^*_n)- Q(P;\widehat{\pi}_n)\big]\geq 0,
\end{align}
where $\mathbb{E}_{P_n}$ is the expectation with respect to the i.i.d.\ sample of observable random variables $(W_i)_{i=1,\dots,n}$ used to estimate $\widehat{\pi}_n$.
The next few subsections build up to a final result providing asymptotic convergence guarantees for $\widehat{\pi}_{n}$ to $\pi^{*}_{n}$ in terms of statistical regret.
\subsection{Assumptions}
We make the following assumptions.
\begin{assumption}[VC-class]\label{ass:VCdimension}
	There exists constants $0\leq\nu<1/2$ and $N\geq1$ such that $\emph{VC}(\Pi_n)\lesssim n^\nu$ for all $n\geq N$.
\end{assumption}Assumption \ref{ass:VCdimension} restricts the policy class to have finite VC-dimension, which is a standard requirement for controlling the complexity of a policy class in the classification literature. The VC-dimension of the policy-class $\Pi$ is defined as the largest interger $m$ such that there exist points $x_1,\dots,x_m$ that are shattered by $\Pi$, i.e.\@ where the policy values $\pi(x_1),\dots,\pi(x_m)$ can take on all $2^m$ possible combinations in $\{0,1\}^m$ (for more on the VC-dimension, see \citealp{wainwright_2019}).
Several practically relevant classes of treatment rules satisfy this requirement, including the linear-index and quadrant rules used in the empirical application of Section \ref{sec:empiricalApplication}. Our assumption allows the VC-dimension of the policy class to grow moderately with the sample size, thus allowing the treatment rule to depend on high-dimensional covariates.
\begin{assumption}[Regularity conditions for data-generating process]\label{ass:Regularity}~ 
\begin{itemize}
\item[(i)] There exist constants $\mathcal{C}_{1,\varphi},\mathcal{C}_{1,\alpha}$ such that for all $\{\widetilde g, \widetilde f\}\in \mathcal{G}\times \mathcal{F}$ \begin{align*}\norm{\varphi_\ell(\widetilde g; X) - \varphi_\ell(g; X)}_{L_\infty(P_X)}&\leq \mathcal{C}_{1,\varphi} \cdot \norm{\widetilde g - g }_{L_\infty(P_V)},\\
\norm{\alpha_\ell(\{\widetilde g, \widetilde f\}; V) - \alpha_\ell(\{g, f\}; V)}_{L_\infty(P_V)}&\leq \mathcal{C}_{1,\alpha} \cdot \left(\norm{\widetilde g - g }_{L_\infty(P_V)} + \norm{\widetilde f - f }_{L_\infty(P_V)} \right),
   \end{align*} for $\ell=0,\dots, L$.
\item[(ii)]There exist constants $\mathcal{C}_{2,\varphi},\mathcal{C}_{2,\alpha}$ such that for all $\{\widetilde g, \widetilde f\}\in \mathcal{G}\times \mathcal{F}$  \begin{align*}\norm{\varphi_\ell(\widetilde g; X) - \varphi_\ell(g; X)}_{L_2(P_V)}&\leq \mathcal{C}_{2,\varphi} \cdot \norm{\widetilde g - g }_{L_2(P_V)},\\
\norm{\alpha_\ell(\{\widetilde g, \widetilde f\}; V) - \alpha_\ell(\{g, f\}; V)}_{L_2(P_V)}&\leq \mathcal{C}_{2,\alpha} \cdot \left(\norm{\widetilde g - g }_{L_2(P_V)} + \norm{\widetilde f - f }_{L_2(P_V)} \right),
   \end{align*}
    for $\ell=0,\dots, L$.
   \item[(iii)]  There exist constants $\mathcal{C}_{3,\varphi},\mathcal{C}_{3,\alpha}$ such that for all $\{\widetilde g, \widetilde f\}\in \mathcal{G}\times \mathcal{F}$ 
   \begin{align*}
   \norm{\varphi_\ell(\widetilde g; X)}_{L_\infty(P_X)}&\leq\mathcal{C}_{3,\varphi},\\
      \norm{\alpha_\ell(\{\widetilde f,\widetilde g\};V)}_{L_\infty(P_V)}&\leq\mathcal{C}_{3,\alpha},
   \end{align*}
   for $\ell=0,\dots,L$.
\item[(iv)]The irreducible noise $\varepsilon_i \coloneqq U_i - g(V_i)$ is a sub-Gaussian vector conditional on $V_i$, with conditional variance ${\rm Var}(\varepsilon_i \mid V_i )= \Sigma(V_i)$ satisfying $\norm{\lambda_{\rm max}(\Sigma(V))}_{L_\infty(P_V)}\leq \overline{\lambda}< \infty$.
\end{itemize}
\end{assumption}
Assumptions 5(i) and 5(ii) impose Lipschitz continuity of the component functions and Riesz-representers with respect to the nuisance component in the $L_\infty$ and $L_2$-norm, respectively. These requirements are typically met under mild conditions within the framework of this paper. For the Balke-Pearl bounds of Example \ref{balkePearl}, these assumptions hold under the overlap condition whenever $\mathcal{G}$ and $\mathcal{F}$ are subsets of the space of bounded functions\footnote{That is, there exists a constant $B>0$ such that $\|\{\widetilde g, \widetilde f\}\|_{L_\infty(P_V)}\leq B, \, \forall \{\widetilde g, \widetilde f\} \in \mathcal{G}\times \mathcal{F} $}, which is automatically satisfied since $U_i=(Y_i, D_i, Z_i)'$ is a vector of random variables with bounded support. Assumption \ref{ass:Regularity}(iii) is a uniform bound on the component functions and Riesz-representers, the former implying uniform boundedness of the scores $\Gamma(g;\cdot)$. Assumption \ref{ass:Regularity}(iv) is a standard requirement in statistical learning theory restricting the tail behaviour of the statistical noise $\varepsilon_i$. It is automatically satisfied when $U_i$ has bounded support, as in the Balke-Pearl bounds, but also allows for outcomes with unbounded support whose conditional distributions have sufficiently thin tails. Together with Assumption \ref{ass:Regularity}(iii), this assumption implies sub-gaussianity of $\Gamma^{\texttt{NO}}(\{g,f\},W_i)$.

The next two assumptions impose requirements on the estimators for the nuisance components.
\begin{assumption}[Regularity conditions for fist-step estimators]\label{ass:UniformConsistency}~ 
\begin{itemize}
\item[(i)] The estimators of the nuisance functions $\{\widehat g_n, \widehat f_n\}$ belong to the function classes $\mathcal{G}\times \mathcal{F}$ with probability 1.
\item[(ii)] There exists a constant $\mathcal{C}_4>0$ such that
\begin{align*}
		\norm{\widehat{g}_n - g}_{L_\infty(P_V)}  \leq \mathcal{C}_4,\\
		\norm{\widehat{f}_n - f}_{L_\infty(P_V)}  \leq \mathcal{C}_4,
\end{align*}
with probability approaching 1 as $n\to \infty$.
\end{itemize}
\end{assumption}
Part (i) of Assumption \ref{ass:UniformConsistency} is needed to ensure the validity of the Lipschitz continuity requirements of Assumption \ref{ass:Regularity} for the component functions and Riesz-representers when evaluated at the first-stage estimates. In the context of the Balke-Pearl bounds, it is satisfied when ${\widehat g_n, \widehat f_n}$ are uniformly bounded and the estimated propensity score $\widehat z(X_i)$ is uniformly bounded away from 0 and 1, with probability one. The first condition is satisfied by virtually any estimation procedure when the outcomes $U_i$ have bounded support. The second requirement can be guaranteed under appropriate trimming of the estimated propensities. Part (ii) requires that estimation errors for the nuisance components are uniformly bounded, which is satisfied under Assumption \ref{ass:UniformConsistency}(i) when $\mathcal{G}\times\mathcal{F}$ is a subset of the space of bounded functions. When $U_i$ has unbounded support and $\mathcal{G}\times\mathcal{F}$  includes unbounded functions, a more primitive condition for (ii) would be uniform consistency of the first stage estimates, that is $\|\{\widehat{g}_n, \widehat f_n \} - \{g,f\}\|_{L_\infty(P_V)}  \rightarrow_p 0$.\footnote{However, it should be noted that the uniform consistency requirement is not completely innocuous when $\{\widehat g_n, \widehat f_n\}$ are machine learning estimators \citep{farrell2020}.}
\begin{assumption}[$L_2$ convergence rates]\label{ass:L2rates}
	The estimators of the nuisance functions satisfy
	\begin{align*}
	&\E_{P_n}\left[ \norm{ \widehat g_n - g }^2_{L_2(P_V)} \right] \leq \frac{r_n}{n^{1/2}},\\ &\E_{P_n}\left[ \norm{ \widehat f_n - f }^2_{L_2(P_V)} \right] \leq \frac{r_n}{n^{1/2}},
	\end{align*}
for some sequence $r_n=o(1)$. 
\end{assumption} The above requirement on the $L_2$-convergence rates for the learners of the nuisance functions is a standard assumption in the semiparametric inference literature (see, e.g., \citeauthor{farrell2015}, \citeyear{farrell2015}, and \citeauthor{chernozhukov2018}, \citeyear{chernozhukov2018}). It can be shown to provably hold for traditional nonparametric estimation methods such as sieve methods (\citeauthor{chen2007}, \citeyear{chen2007}) as well as modern black-box machine learning algorithms including Lasso (see, e.g., \citeauthor{farrell2015}, \citeyear{farrell2015}), deep neural networks (\citeauthor{farrell2020}, \citeyear{farrell2020}), boosting and others, for which stronger guarantees such as Donsker-type properties are typically not available.
The ability to invoke a mild $L_2$-convergence requirement is a virtue of the combined use of Neyman-orthogonalization and sample-splitting, a key insight brought forward by \cite{chernozhukov2018} for semiparametric GMM inference, and subsequently leveraged by \cite{atheywager2021} and \cite{fosterSyrgkanis} in the context of statistical learning problems.\footnote{Unlike \cite{atheywager2021}, our assumptions do not allow to trade-off accuracy in the estimation across the different nuisance functions. This is because our framework allows for $\varphi_\ell(g;x)$ to be a potentially non-linear functional of the nuisance functions $g$, as is the case in Examples \ref{manskiBounds}-\ref{ManskiPepper}, thus precluding such double-robustness property.}

Finally, we present an assumption that concerns the distribution of the component functions $\varphi_\ell$ at the population level.
\begin{assumption}[Margin]\label{ass:Margin}There exist constants $\mathcal{C}_m>0$ and $\gamma\geq0$ such that
		\begin{align*}
		\mathbb{P}_X\Big( 0<|\varphi_\ell(g;X)|\leq t\ \Big)  \leq \mathcal{C}_m t^\gamma , \quad \forall \,t>0.	\end{align*}
		for $\ell=1,\dots , L$.
\end{assumption}
The above assumption restricts the extent to which the distribution of the component functions $\varphi_\ell(g;X)$ can concentrate around the point of non-differentiability 0 and it is a form of ``margin assumption", first introduced by \citet{mammenTsybakov1999}. Such an assumption has been widely used in statistics to obtain fast learning rates in classification problems (see, e.g., \citeauthor{arlotBartlett2011}, \citeyear{arlotBartlett2011}). 
Notice that the above formulation for the margin assumption restricts the concentration of probability for the distribution of the components functions in a neighbourhood of 0, but still allows for arbitrary probability mass at 0.
\begin{example}[$\gamma=1$]\label{exampleBoundedDerivative}
	Suppose $X$ contains an absolutely continuous covariate $\mathrm{\tilde{x}}$ and $\varphi_\ell(g;X)\cdot\mathbbm{1}\{\mathrm{\tilde{x}}\neq 0\}$ is absolutely continuous with density bounded above by $\overline{f}$ for $\ell=1,\dots,L$. Then Assumption (\ref{ass:Margin}) holds with $\gamma=1$ and $\mathcal{C}_m= 2\overline{f} $.
\end{example} 
 \begin{example}[$\gamma=\infty$]\label{exampleFullMargin}
 	Suppose there exists a $t_0>0$ such that $\mathbb{P}_X\left( 0<|\varphi_\ell(X)|<t_0 \right) = 0 $ for $\ell=1,\dots,L$. Then Assumption (\ref{ass:Margin}) holds with $\gamma=\infty$ and some $\mathcal{C}_m>0$.
 \end{example}
In the context of the Balke-Pearl bounds from Example \ref{balkePearl} with resolution of ambiguity via Minimax Regret, Assumption \ref{ass:Margin} restricts the extent to which the CATE bounds $\overline \tau,\underline \tau$ can concentrate around 0 in the data-generating process. Under $\gamma=\infty$ the support of each CATE bound is required to be fully separated from 0, while $\gamma=1$ requires that each CATE bound has bounded density in a neighborhood of 0.
 
 
In the next section we present our theoretical results based on Assumptions \ref{ass:VCdimension}-\ref{ass:Margin}.
\subsection{Regret convergence rates}\label{subsec:RegretConvergenceRates}
In this section, we provide asymptotic rates of convergence for the regret of the estimated policy $R_{n}(P;\widehat{\pi}_{n})$ as defined in \eqref{regret}. In line with the existing literature, we study \textit{uniform} regret bounds that are valid for all distributions $P\in\mathcal{P}$ satisfying Assumptions \ref{ass:VCdimension}-\ref{ass:Margin}. All results in this section are thus intended to hold uniformly in the above sense, and we will drop the dependence on $P$ for notational convenience.

We begin by noticing that controlling the convergence of $\widehat{\pi}_n$ to the best-in-class policy $\pi^*_n$ intuitively requires accounting for: 1) the estimation error in the component functions $\varphi_\ell$ and influence function adjustments $\phi_\ell$ due to estimation of the nuisance components $\{g,f\}$, 2) the difference between the population Neyman-orthogonal score and true score\footnote{That is, we need to account for the fact that we have added the influence function adjustments to the component functions.}, and  3) the fact that we estimate our policy using a sample from the distribution of the covariates $X_i$ rather than their true distribution. We define the following quantities:
\begin{align*}
	\widehat{Q}_n^{\texttt{NO}}(\pi)& = \frac{1}{n}\sum_{i=1}^n \big(2{\pi}(X_i) -1\big)\cdot \Gamma^{\texttt{NO}}(\{\widehat g, \widehat f\};W_i),\\
	{Q}^{\texttt{NO}}_n(\pi)& = \frac{1}{n}\sum_{i=1}^n \big(2{\pi}(X_i) -1\big)\cdot\Gamma^{\texttt{NO}}(\{g,f\},W_i),
\end{align*}
and formalize this intuition in the next proposition.
\begin{proposition}\label{prop:regretERMbound} The regret of $\widehat{\pi}_n$ obeys the following bound:
\begin{align}\label{eq:regretERMbound}
	R_n(\widehat{\pi}_n) \leq &2\mathbb{E}\left[\sup_{\pi \in \Pi_n}\left|\widehat{Q}^{\emph{\texttt{NO}}}_n(\pi) - Q^{\emph{\texttt{NO}}}_n(\pi)\right|\right] +\mathbb{E}\left[\sup_{\pi \in \Pi_n}\left|Q^{\emph{\texttt{NO}}}_n(\pi) - Q(\pi)\right|\right].
\end{align}
\end{proposition}

The second term in the above bound accounts for points 2) and 3). $Q_n(\pi) - Q(\pi)$ is a centred (mean-zero) empirical process and therefore its uniform expectation can be shown to be $O\left(\sqrt{\text{VC}(\Pi_n)/n}\right)$ using symmetrization and chaining arguments \citep[see, e.g.,][Ch. 5.3]{wainwright_2019}. 
Controlling the first term, which accounts for point 1), is particularly challenging and requires tailored arguments that deal with the particular form of the population scores in \eqref{functionalFormScores}, in particular their lack of full differentiability.
	\begin{lemma}\label{mainLemma}
	Suppose that Assumptions \ref{ass:VCdimension}-\ref{ass:Margin} hold and define $\kappa_n=\lfloor n(1-1/K) \rfloor$. Then we have
	\begin{align*}
		\mathbb{E}_{P_n}\left[\sup_{\pi \in \Pi_n}\left|\widehat{Q}^{\emph{\texttt{NO}}}_n(\pi) - Q^{\emph{\texttt{NO}}}_n(\pi)\right|\right]= O\left( \frac{r_{\kappa_n}}{\sqrt{n}} + \sqrt{\frac{\emph{VC}(\Pi_n)}{n}}  +  \left( \frac{r_{\kappa_n}}{\sqrt{n}}\right) ^{\frac{\gamma+1}{\gamma+2}}\right).
	\end{align*}
\end{lemma}Lemma \ref{mainLemma} is the central result of this paper. It provides an asymptotic rate of convergence to zero of the empirical process $\left|\widehat{Q}^{\texttt{NO}}_n(\pi) - Q^{\texttt{NO}}_n(\pi)\right|$ uniformly over the policy class $\Pi_n$, which depends on the VC-dimension of the class and the degree of concentration of the component functions $\varphi_\ell(g;X)$ around 0, as indexed by $\gamma$. 
 In order to convey intuition on this result we provide a brief outline of the proof, which is based on the decomposition 
\begin{equation}\label{eq:DecompositionQhat-Qtilde}
\begin{aligned}
	\widehat{Q}^{\texttt{NO}}_n(\pi) - Q^{\texttt{NO}}_n(\pi) &=\frac{1}{n}\sum_{i=1}^n  (2\pi(X_i)-1)\cdot\left[\widehat{\Gamma}^{\texttt{NO}}\left(\{\widehat g^{-k(i)},\widehat f^{-k(i)}\},W_i\right) -\Gamma^{\texttt{NO}}\left(\{ g,f\},W_i\right)\right] \\ &=A_0(\pi) + \sum_{\ell=1}^L a_\ell\cdot \left[A_{1,\ell}(\pi) + A_{2,\ell}(\pi) + A_{3,\ell}(\pi)\right],
\end{aligned}
\end{equation}
where
\begin{align*}
&A_0(\pi)=\frac{1}{n}\sum_{i=1}^n (2\pi(X_i)-1) \cdot\left[\varphi_0^{\texttt{NO}}(\{\widehat g^{-k(i)}, \widehat f^{-k(i)}\},W_i) - \varphi_0^{\texttt{NO}}(\{ g, f\},W_i)\right],\\
&A_{1,\ell}(\pi)= \frac{1}{n}\sum_{i=1}^n (2\pi(X_i)-1) \cdot\Big[\varphi_\ell^{\texttt{NO}}(\{\widehat g^{-k(i)}, \widehat f^{-k(i)}\},W_i) - \varphi_\ell^{\texttt{NO}}(\{ g, f\},W_i)\Big]\cdot\mathbbm{1}\left\{ \varphi_\ell\left(\widehat g^{-k(i)};X_i\right) >0 \right\},\\
&A_{2,\ell}(\pi)=\frac{1}{n}\sum_{i=1}^n (2\pi(X_i)-1)\cdot \phi_{\ell}(\{g,f\};W_i) \cdot\Big[ \mathbbm{1}\left\{\varphi_\ell(\widehat g^{-k(i)};X_i) \geq0\right\} - \mathbbm{1}\left\{\varphi_\ell(g;X_i) \geq0  \right\} \Big],\\
&A_{3,\ell}(\pi)=\frac{1}{n}\sum_{i=1}^n (2\pi(X_i)-1)\cdot  \varphi_{\ell}(g;X_i) \cdot\Big[ \mathbbm{1}\left\{\varphi_\ell(\widehat g^{-k(i)};X_i)  \geq0\right\} - \mathbbm{1}\left\{\varphi_\ell(g;X_i)\geq0  \right\} \Big].
\end{align*}
Terms $A_0(\pi)$ and $A_{1,\ell}(\pi)$ can be controlled using similar arguments to \citet{atheywager2021} and are responsible for the $O\left(r_{\kappa_n}/\sqrt{n}\right)$ term in the bound of Lemma \ref{mainLemma}. The de-biasing properties of Neyman-orthogonalization combined with sample-splitting play a crucial role in this context, as they ensure that the error in estimating $\varphi_\ell(g;x)$ only has a second-order contribution. As a result, term $A_{1,\ell}(\pi)$ scales with the mean-squared estimation error in the nuisance functions and, under Assumption \ref{ass:L2rates}, its expectation decays faster than $1/\sqrt{n}$ uniformly over $\Pi_n$.\footnote{Notice that, by virtue of sample-splitting, the presence of the indicator $\mathbbm{1}\{\varphi_\ell(\widehat g^{-k(i)};X_i)\geq0 \}$ is immaterial when controlling the expectation of $A_{1,\ell}(\pi)$ uniformly over $\Pi_n$.} If plug-in (non-orthogonalized) estimates for $\varphi_\ell$ are instead used to form the score estimates $\widehat{\Gamma}_i$, the estimation error in the nuisance functions  has a first-order impact on term $A_{1,\ell}(\pi)$. As a result, its uniform expectation would scale with the $L_1$ estimation error which, under Assumption \ref{ass:L2rates}, implies the much slower convergence $\E[\sup_{\pi \in \Pi_n}|A_{1,\ell
}(\pi)|]= o(n^{1/4})$. 

For term $A_{2,\ell}(\pi)$, the mean-zero property of the influence function adjustments together with sample-splitting ensures that this term is a centred empirical process and thus it is responsible for a $O\left(\sqrt{\text{VC}(\Pi_n)/n}\right)$ contribution again by symmetrization and chaining arguments. 

Finally, for term $A_{3,\ell}(\pi)$ we show that 
\begin{align*}
\E\left[\sup_{\pi\in\Pi_n}A_{3,\ell}(\pi)\right]\leq \E\Bigg[\left|\varphi_\ell(g;X_i) \cdot \left( \mathbbm{1}\left\{\varphi_\ell^{-k(i)}(\widehat g^{-k(i)};X_i) \geq0\right\} - \mathbbm{1}\left\{\varphi_\ell(g;X_i) \geq0  \right\} \right)\right|\Bigg],
\end{align*}
where the RHS can be recognized to be the classification loss of an estimator for the sign of $\varphi_\ell(g;x)$ based on thresholding  $\varphi_\ell(\widehat g^{-k(i)};x)$. Rates of convergence in binary classification problems intuitively depend on the degree of separation of the true regression function from 0, as indexed by $\gamma$. We thus leverage results from the literature on classification \citep{audibertTsybakov2007} to quantify the contribution of $A_{3,\ell}$ in the bound of Lemma \ref{mainLemma} in terms of $\gamma$.

We are now ready to combine the rates of convergence for the three terms in Proposition \ref{prop:regretERMbound} to obtain a final regret bound for our proposed estimation procedure.\begin{theorem}\label{mainTheorem}
	Suppose Assumptions \ref{ass:Regularity}-\ref{ass:Margin} hold. Then the regret obeys
	\begin{align*}
		R_n(\widehat{\pi}_n)=O\left( \sqrt{\frac{\emph{VC}(\Pi_n)}{n} } \vee \left( \frac{r_{\kappa_n}}{\sqrt{n}}\right) ^{\frac{\gamma+1}{\gamma+2}}\right). 
	\end{align*}
\end{theorem}
We see that regret convergence for our policy learning procedure happens at a rate corresponding to whichever is the leading term in the asymptotic expansion of Lemma \ref{mainLemma}, which depends on $\nu$ and $\gamma$. When the policy class $\Pi_n$ has fixed VC-dimension ($\nu=0$), regret convergence happens at rates ranging from $o(n^{1/4})$ in the least favourable case ($\gamma=0$) to $O(\sqrt{\text{VC}(\Pi)/n})$ in the most favourable case ($\gamma=\infty$). The latter case is in line with existing results for policy learning with point-identified CATE, in which full-differentiability of the scores leads to $\sqrt{\text{VC}(\Pi_n)/n}$ learning rates \citep[see][]{kitagawatetenov2018, atheywager2021,fosterSyrgkanis}. For the intermediate case $\gamma=1$ of Example \ref{exampleBoundedDerivative} our procedure guarantees regret convergence at rate $o(n^{1/3})$.

It is useful to compare the performance guarantees in this paper with \citet{Pu_2021}, whose procedure involves the use of non-orthogonalized estimates for the scores with sample-splitting. 
They show that the regret of a policy estimated via the maximization $(\ref{estimationPolicy})$ based on cross-fitted non-orthogonalized scores is upper bounded by the $L_1$-norm of the estimation error in the nuisance functions. Under Assumption \ref{ass:L2rates}, this implies $o(n^{1/4})$ convergence for the regret, which is strictly slower than our rates \textit{for all values} of $\gamma>0$. The faster speed of convergence guaranteed by our procedure is not just due to a refined proof strategy but crucially depends on the use of Neyman-orthogonalization, as elucidated by our discussion of Lemma \ref{mainLemma}.\begin{remark}The procedure of \citet{Pu_2021} also differs from ours in its final implementation, which in their case is carried out via support vector machines (SVM) with $\Pi_n$ assumed to be a reproducing kernel Hilbert space. While the use of surrogate losses (such as the hinge loss in SVM) to convexify problem (\ref{estimationPolicy}) can bring considerable computational benefits in terms of speed and scalability, it comes at the cost of even slower convergence guarantees than the $o(n^{-1/4})$ discussed above. We stress that our insights regarding the benefits of Neyman-orthogonalization in terms of faster learning rates apply irrespective of the final implementation. Notice also that the use of surrogate loss functions does not guarantee convergence of the estimated optimal policy to the best-in-class $\pi^*_n$ in general when the policy class $\Pi_n$ does not contain the ``first-best" policy $\mathbbm{1}\{\Gamma(g;x)\geq 0\}$, as shown by the recent work of \cite{kitagawaSakaguchi2021}.  \end{remark}

\section{Empirical application}\label{sec:empiricalApplication}
In this section we apply the methods discussed in this paper to data from the National Job Training Partnership Act (JTPA) Study. This study randomly selected applicants to receive various training and services, including job-search assistance, for a period of 18 months. The study collected background information on applicants before random assignment and then recorded their earnings in the 30-month period following treatment assignment.  
\citet{kitagawatetenov2018} apply their EWM method to a sample of 9,223 adult JTPA applicants to estimate the optimal allocation of \textit{eligibility} into the programme that maximizes individual earnings across the population. In particular, they take total individual earnings in the 30 months after assignment as the welfare outcome measure $Y_i$, and consider policies that allocate eligibility in the programme based on the individual's observable characteristics. \citeauthor{kitagawatetenov2018}'s analysis is from an \textit{intent-to-treat} perspective as they focus on the problem of deciding who should be given eligibility to participate in the programme. Since eligibility in the JTPA study is randomly assigned, the effect of eligibility on earnings is point identified from the data and methods for policy learning under point-identification can be applied in this setting.
We depart from \citet{kitagawatetenov2018} and instead consider optimal assignment of \textit{actual participation} in the training. This analysis would be of interest to a policy-maker that expects to achieve (close to) perfect compliance to her treatment decision, e.g. when participation is made a condition for receipt of a  generous unemployment benefit.\footnote{No financial incentive had been put in place to promote compliance in the implementation of the JTPA study.} Compliance in the JTPA study is imperfect as roughly 23\% of applicants' participation status $D_i=0,1$ deviates from their assigned eligibility status $Z_i=0,1$, as shown in Table \ref{tableEligibility}. As a result, random assignment of the eligibility instrument $Z_i$ is not sufficient to point-identify the effect of participation in the training, motivating the use of the methods proposed in this paper. 
\begin{table}[ht]
	\centering
	\caption{Joint distribution of eligibility and participation, JTPA study}
	\label{tableEligibility}
	\begin{threeparttable}
	\begin{tabular}[t]{cccc}
		\toprule
		&\multicolumn{2}{l}{Eligibility ($Z_i$) }\\
		
		Participation ($D_i$)&0&1&Total\\
\midrule
		0	&3047&2118&5165\\
		1	&43&4015&4058\\
		Total		&3090&6133&9233\\
		\bottomrule
	\end{tabular}
\begin{tablenotes}
	\footnotesize 
	\item Data source: \citet{kitagawatetenov2018} and \citet*{abadieAngrist2002}.
\end{tablenotes}
\end{threeparttable}
\end{table}%

For partial identification of the CATE we consider the Balke-Pearl scheme of Example \ref{balkePearl}, where bounds for the 30-month post-treatment earnings are $Y_L=\$0$ and $Y_U=\$59,640$.\footnote{The outcome upper bound corresponds to the 97.5th percentile of the earnings distribution rather than highest recorded value of $\$155,760$. Outcome bounds in Balke-Pearl bounds effectively impute unidentified expected earnings for never-takers and always-takers. Restricting expected earnings to be below such high quantile is in effect a mild requirement which brings considerable identification power.} We compare this with point-identification of the CATE as the conditional local average treatment effect (LATE), predicated under the assumption of no unobserved heterogeneity. We subtract \$1216 from both the CATE bounds and the conditional LATE; this is the average cost of services per actual treatment, estimated from Table 5 in \cite{bloom1997}. Following \citet{kitagawatetenov2018}, we condition treatment assignment on two pre-treatment variables: the individual's years of education and earnings in the year prior to assignment.
Estimation of the optimal policy follows the procedure described in Section \ref{sec:Estimation}, with $K=10$ evenly-sized data folds used to form cross-fitted Neyman-orthogonal estimates for the CATE bounds and conditional LATE functions.
The nuisance functions are estimated via boosted regression trees, performed by the MATLAB function \verb+fitrensenmble+.\footnote{Tuning parameters have been chosen via cross-validation within each data-fold. For further details on the estimation procedure we refer to the MATLAB documentation for the command.}

Figure 1 demonstrates cross-fitted plug-in estimates for the CATE bounds (a) and the LATE/minimax regret scores (b), where the size of the dots indicates the number of individuals with different covariate values. 
We first notice that the estimated CATE lower bounds are negative for the whole sample, and thus the maximin impact optimal policy never assigns treatment in this application. We therefore focus our analysis on minimax regret (MMR).\footnote{Maximin welfare also results in no treatment for the whole population in this application.} Comparison of the conditional LATE and MMR scores highlights how partial identification leads to increased variation of the scores across different levels of education and pre-program earnings. In particular, the MMR scores are considerably higher (and positive) for individuals with fewer years of education and smaller pre-programmes. The conditional LATE estimates display less overall variation over the support of the covariates, compared to the MMR scores, but are lower for individuals with 0 pre-programme earnings.


\begin{figure}\label{boundsScatter}
\centering
\label{blala}
\begin{subfigure}[b]{1\linewidth}
\includegraphics[width=\linewidth]{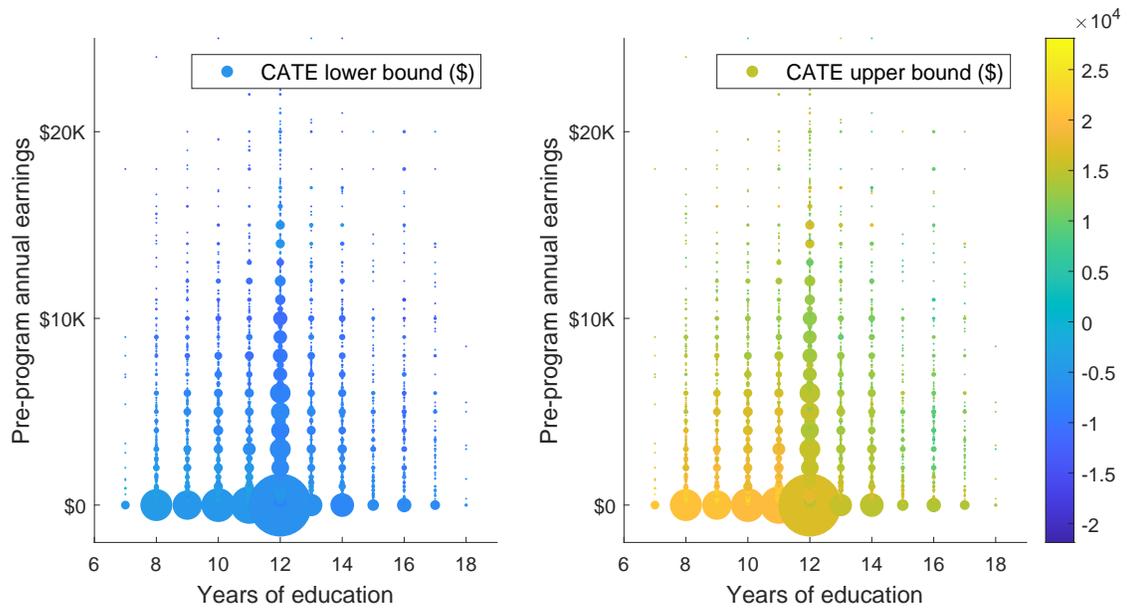}
\caption{CATE bounds}
\end{subfigure}

\begin{subfigure}[b]{1\linewidth}
\includegraphics[width=\linewidth]{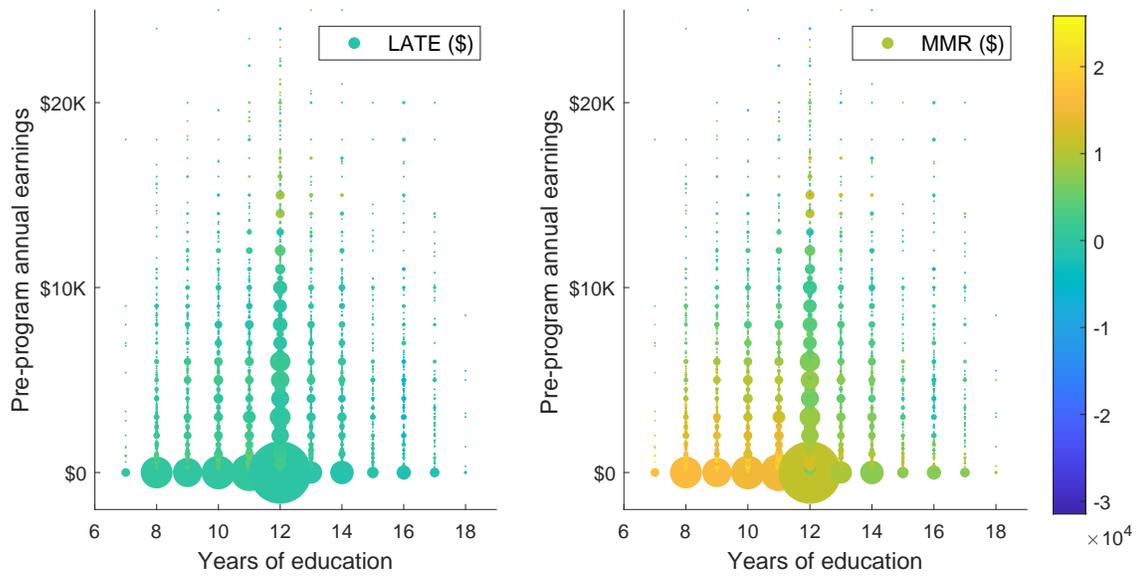}
\caption{Scores}\label{fig:gull}
\end{subfigure}
\caption{JTPA -- Plug-in cross-fitted estimates (net of \$1216)}
\label{fig:animals}
\end{figure}

\begin{figure}[H]
	\centering
	\includegraphics[keepaspectratio,scale=0.8]{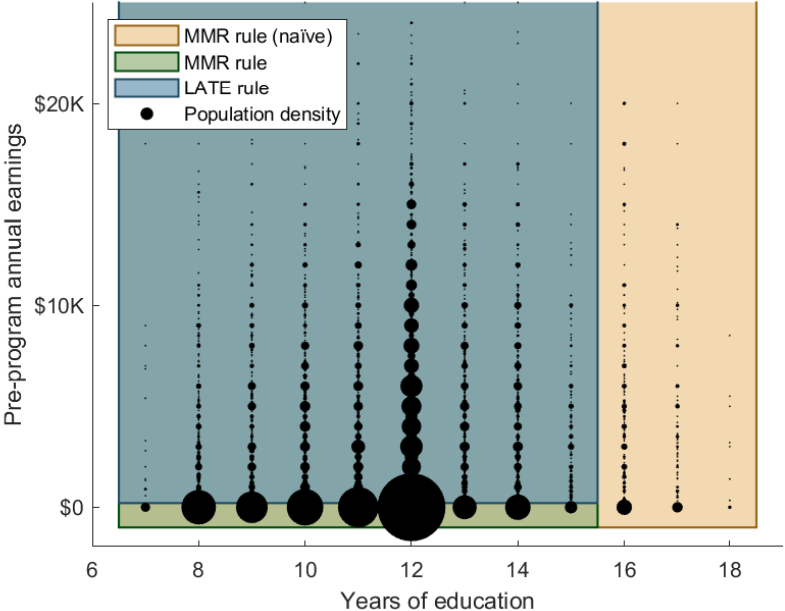}
	\caption{Estimated optimal policies from the quadrant policy class conditioning on years of education and pre-programme earnings.}\label{quadrantGraph}
\end{figure}

We consider three alternative choices for the candidate policy class $\Pi$. The first is the class of quadrant treatment policies. To be assigned to treatment according to this policy, an individual's education and pre-program earnings have to be above (or below) some specific threshold. Figure \ref{quadrantGraph} illustrates the optimal quadrant treatment policies based on Neyman-orthogonalized cross-fitted MMR and LATE scores, where the colored shaded areas indicate individuals required to undertake job training by the respective policies. The optimal MMR policy (green) assigns treatment to individuals with education below 15 years and pre-treatment earnings below \$39,952. The optimal LATE policy (blue) selects the same threshold for education, but selects individuals with pre-treatment earnings above \$200 for treatment.
While the two policies appear similar, they substantially differ in the proportion of population assigned to treatment (96\% by the MMR policy versus 64\% by the LATE policy), as shown in Table \ref{table:empirical findings}. This is due to the large concentration of individuals with pre-treatment earnings close to (or equal) zero. As a result, 32\% of individuals receive a different treatment assignment across the two policies. Figure \ref{quadrantGraph} also shows the optimal ``na\"ive" MMR policy based on cross-fitted but non-orthogonalized scores (yellow), which recommends participation into the programme for the entire population. 
{\small
\begin{table}[H]
	\centering
	\caption{Treatment proportions of alternative treatment assignment policies}
	\label{table:empirical findings}
	\scalebox{1}{
		\begin{threeparttable}
			
			\begin{tabular}{lccccccccccc}
				\toprule
				&&&\shortstack{Share of Population\\ to be treated} & \multicolumn{3}{@{}c}{\shortstack{Share of Population receiving\\ same treatment as}}\\
		\cmidrule(l{1pt}r{1pt}){5-7}	&&&&MMR (na\"ive) &MMR & LATE \\
				\hline \hline
				\multicolumn{4}{@{}c}{\textbf{Quadrant Rule}}\\
				Minimax Regret (na\"ive) &&&$1.00$&--& &\\
				Minimax Regret&&&$0.96$&$0.96$&--&\\
				LATE&&&$0.64$&$0.68$&$0.68$&-- \\ \midrule
				
				\multicolumn{4}{@{}c}{\textbf{Linear Index Rule}}\\
				Minimax Regret (na\"ive) &&&$0.99$&--&&\\
				Minimax Regret&&&$0.96$&$0.96$&--&\\
				LATE&&&$0.69$& $0.69$&$0.70$&--\\ \midrule
				
				\multicolumn{4}{@{}c}{\textbf{Linear Index Rule + ${\rm \bf edu^2}+{\rm \bf edu^3}$}}\\
				Minimax Regret (na\"ive) &&&$0.99$&--&&\\
				Minimax Regret&&&$0.96$&$0.97$&--&\\
				LATE&&&$0.75$& $0.75$&$0.75$&--\\
				\bottomrule
			\end{tabular}
			\begin{tablenotes}
				\footnotesize 
				\item The rows labeled ``Minimax Regret (na\"ive)" give information on the estimated optimal minimax regret policy based on the scores in Equation \eqref{scoresMMR} with the Balke-Pearl CATE bounds of Example \ref{balkePearl}, without Neyman-orthogonalization. The rows labeled ``Minimax Regret" give information on the estimated optimal minimax regret policy with Neyman-orthogonalization. The rows labeled ``LATE" give information on optimal policy for Neyman-orthogonal scores for the conditional LATE.
			\end{tablenotes}
	\end{threeparttable}}
\end{table}
}

\begin{figure}[H]
	\centering
	\includegraphics[keepaspectratio,scale=0.8]{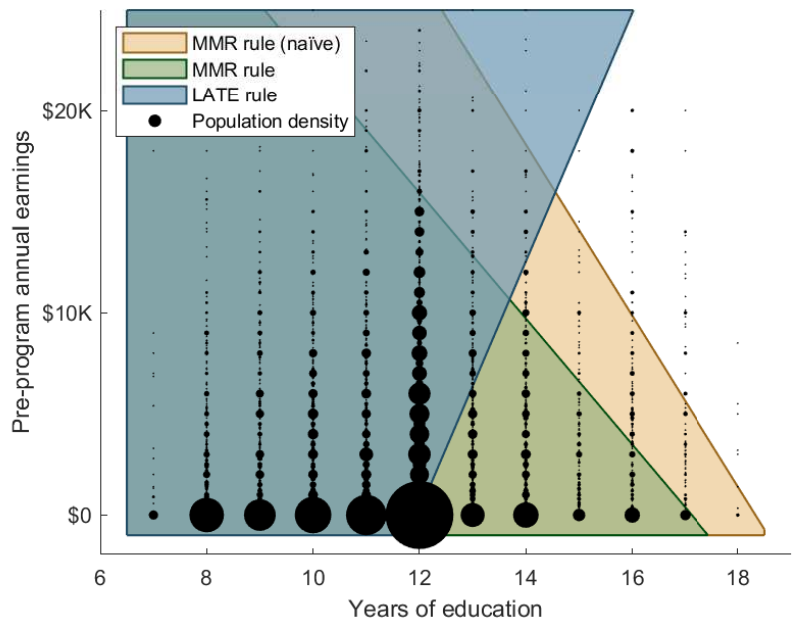}
	\caption{Estimated optimal policies from the linear-index policy class}\label{linearGraph}
\end{figure}

Second, we consider the class of linear treatment policies. This class consists of policies that assign treatment to an individual according to whether a linear index in his observable characteristics is above a certain threshold. Figure \ref{linearGraph} illustrates how the direction of treatment assignment as a function of prior earnings differs between the MMR and LATE policy in a similar fashion to the quadrant rules; contrary to the LATE policy, MMR prioritizes treatment assignment to individuals with lower pre-program earnings. Nonetheless, 70\% of the population still receives the same treatment under the two different policies, in light of the relatively low concentration of individuals in the areas of the covariate space where the two policies differ. Similarly to the quadrant policy rule, the MMR policy assigns treatment to a larger share of the population (95\%) compared to the LATE policy (69\%). The na\"ive MMR policy is qualitatively similar to the one using Neyman-orthogonolization, but recommends programme participation to a larger share of individuals.

\begin{figure}[H]
	\centering
	\includegraphics[keepaspectratio,scale=0.8]{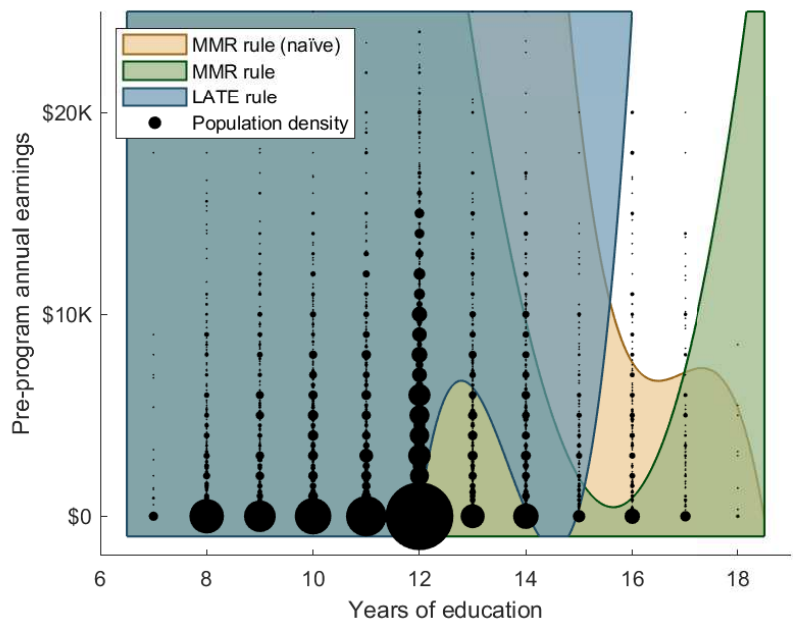}
	\caption{Estimated optimal policies from the linear-index policy class conditioning on years of education, ${\rm (education)^2}$, ${\rm (education)^3}$, and pre-programme earnings.}\label{linearGraph3}
\end{figure}

Finally, we consider linear treatment policies that additionally include quadratic and cubic terms for education. Figure \ref{linearGraph3} shows how the additional flexibility in the policy class leads to rules that are less interpretable but maintain similar qualitative features compared to the more parsimonious classes previously considered.


\section{Conclusion}\label{sec:Conclusion}
This paper develops a general policy learning framework for estimation of individualized treatment rules when treatment effects are partially identified. By drawing connections between the treatment assignment problem and classical decision theory, we have characterized several notions of optimal treatment policies in the presence of partial identification.
We have shown how partial identification leads to a new policy learning problem where the risk is only directionally-differentiable with respect to a nuisance infinite-dimensional component. We have proposed an estimation procedure that ensures Neyman-orthogonality with respect to the nuisance components and we have provide statistical guarantees that depend on the amount of concentration around the points of non-differentiability in the data-generating process. Our proposed methods are illustrated with an application to the Job Training Partnership Act study, where we have shown that allowing for partial identification delivers substantially different programme participation policies compared to existing methods that assume point-identification. 

There are several avenues for future research. First, it would be interesting to extend the theory of this paper to partial identification via instrumental variables with continuous support.
Second, it would be useful to extend the methods to more general identification sets that incorporate smoothness restrictions on unobserved counterfactual quantities, such as those considered in \cite{simonLeeBounds2018}. Finally, it would be interesting to assess the optimality of our proposed estimation procedure by deriving minimax lower bound rates for semiparametric statistical learning problems with directionally-differentiable risk. 

\newpage
\begin{appendices}\label{Appendix}
\section{Extension to nested min/max operators}\label{appendixExtensionMinMax}
In this section we describe how the proposed estimation procedure and the theoretical results of Section \ref{sec:Estimation} can be extended to scores $\Gamma(g;X)$ that feature nested linear combinations of $\min/\max$ operators. This extension comprises $\min/\max$ operators over multiple components, since $\max\{a,b,c\}=\max\{\max\{a,b\},c\}$.

We begin by noticing that our proposed estimation described in Section \ref{sec:Estimation} can be also defined as follows. First, for each $\min\{a(g;x),b(g;x)\}$ (or $\max$) operator contained in $\Gamma(g;x)$, one substitutes the operator with $a(g;x)$ or $b(g;x)$ based on their cross-fitted plug-in (non-orthogonalized) estimates $\widehat a_i:=a(\widehat g^{-k(i)};X_i)$ and  $\widehat b_i:=b(\widehat g^{-k(i)};X_i)$. Then, the selected component is estimated ($a(g;x)$ or $b(g;x)$) is estimated by their cross-fitted Neyman-orthogonal analogue ($\widehat a^{\texttt{NO}}_i$ or $\widehat b^{\texttt{NO}}_i$). In the presence of nested $\min/\max$ operators, our estimation is generalized as follows. First, in succession from the most inner to the most outer $\min/\max$, each operator is substituted with their smallest/largest argument based on cross-fitted non-orthogonalized estimates. Then, the selected components are estimated by their cross fitted Neyman-orthogonal analogue. As an illustration, consider the hypothetical score
\begin{align*}
\Gamma(g;x) &= d(g;x) + \max\{ c(g;x) + \min\{a(g;x),b(g;x)\} , 0\}\\
 & = d(g;x) + \max\{\underbrace{c(g;x) +b(g;x) +(a(g;x) - b(g;x)) \cdot \mathbbm{1}\{a(g;x) - b(g;x) \leq 0\}}_{\varrho(g;x)}  ,0\}\\
 &=d(g;x) +\varrho(g;x)\cdot \mathbbm{1}\left\{\varrho(g;x)\geq 0\right\}.
\end{align*}
Applying the above procedure to this example gives the following expression for the estimated Neyman-orthogonal score:
\begin{align*}
\Gamma^{\texttt{NO}}(\{\widehat g^{-k(i)},\widehat f^{-k(i)}\},W_i)= \widehat d^{\texttt{NO}}_i + \left[ \widehat{c}^{\texttt{NO}}_i +\widehat{b}^{\texttt{NO}}_i + ( \widehat{a}^{\texttt{NO}}_i - \widehat{b}^{\texttt{NO}}_i)\cdot \mathbbm{1}\{ \widehat a_i - \widehat{b}_i\leq 0\} \right]\cdot \mathbbm{1}\left\{\widehat \varrho_i \geq 0 \right\},
\end{align*}
where
\begin{align*}
\widehat \varrho_i = \widehat{c}_i +\widehat{b}_i + ( \widehat{a}_i - \widehat{b}_i)\cdot \mathbbm{1}\{ \widehat a_i - \widehat{b}_i\leq 0\}.
\end{align*}
We will now show how the theoretical results of Section \ref{sec:Theory} can be generalized to this example.\footnote{ The extension to general scores containing an arbitrary finite number of nested $\min/\max$ operators follows immediately from our discussion of this example.} Following the arguments of Section \ref{subsec:RegretConvergenceRates}, we have
 \begin{align*}
 \widehat Q_n^{\texttt{NO}}(\pi) - Q_n^{\texttt{NO}}(\pi)&= \sum_{i=1}^n(2\pi(X_i)-1)\cdot (\widehat{d}^{\texttt{NO}}_i - d^{\texttt{NO}}_i)\\
&+ \sum_{i=1}^n(2\pi(X_i)-1)\cdot (\widehat c^{\texttt{NO}}_i +\widehat b^{\texttt{NO}}_i  - c^{\texttt{NO}}_i -b^{\texttt{NO}}_i  ) \cdot \mathbbm{1}\{\widehat \varrho_i \geq 0 \} \\
&+ \sum_{i=1}^n(2\pi(X_i)-1)\cdot (\widehat a^{\texttt{NO}}_i  - \widehat b^{\texttt{NO}}_i ) \cdot \mathbbm{1}\{\widehat a_i - \widehat b_i \leq 0 \} \cdot \mathbbm{1}\{ \widehat \varrho_i \geq 0 \} \\
&+ \sum_{i=1}^n(2\pi(x)-1)\cdot ( a^{\texttt{NO}}_i  -  b^{\texttt{NO}}_i) \cdot \left[\mathbbm{1}\{\widehat a_i - \widehat b_i \leq 0 \} -\mathbbm{1}\{ a_i - b_i \leq 0 \} \right]\cdot \mathbbm{1}\{ \widehat \varrho_i \geq 0 \} \\
&+ \sum_{i=1}^n(2\pi(x)-1)\cdot \varrho^{\texttt{NO}}_i \cdot \left[\mathbbm{1}\{\widehat \varrho_i \geq 0 \} -\mathbbm{1}\{ \varrho_i \geq 0 \}\right].
 \end{align*}
 The first term in the expansion has the same structure as $A_{0,\ell}$ and thus obeys the same bound. The second and third term obey the same bound as $A_{1,\ell}$ since, by virtue of sample-splitting, the indicators $\mathbbm{1}\{\widehat \varrho_i \geq 0\}$  and $\mathbbm{1}\{\widehat a_i - \widehat{b}_i \geq 0\}$ are immaterial when controlling the expectation of these term uniformly over $\Pi_n$ (see arguments in the Proof of Lemma 1). The fourth term has the same structure as $A_{2,\ell} + A_{3,\ell}$ except for the presence of the indicator $\mathbbm{1}\{\widehat \varrho_i \geq 0\}$, which again can be shown to be immaterial for controlling the $A_{2,\ell}$-like term by virtue of sample-splitting. For the $A_{3,\ell}$-like term we instead have the bound \begin{align*}
\E&\left[\sup_{\pi\in\Pi_n}\frac{1}{n}\sum_{i=1}^n(2\pi(x)-1)\cdot (c_i-b_i) \cdot \left(\mathbbm{1}\{\widehat c_i - \widehat b_i \leq 0 \} -\mathbbm{1}\{ c_i - b_i \leq 0 \} \right)\cdot \mathbbm{1}\{ \widehat \varrho_i \geq 0 \}\right]\\ &\leq \E\Bigg[\left|(c_i-b_i) \cdot \left(\mathbbm{1}\{\widehat c_i - \widehat b_i \leq 0 \} -\mathbbm{1}\{ c_i - b_i \leq 0 \} \right)\right|\cdot\left|\mathbbm{1}\{ \widehat \varrho_i \geq 0 \}\right|\Bigg]\\
&\leq \E\Bigg[\left|(c_i-b_i) \cdot \left(\mathbbm{1}\{\widehat c_i - \widehat b_i \leq 0 \} -\mathbbm{1}\{ c_i - b_i \leq 0 \} \right)\right|\Bigg],
\end{align*}
which can be bounded in the same fashion as $A_{3,\ell}$ under a margin assumption on $a(g;X)- b(g;X)$. Finally, the fifth term also has the same structure as $A_{2,\ell} + A_{2,\ell}$, and can be controlled using arguments from Section \ref{subsec:RegretConvergenceRates} under a margin assumption on $\rho(g;X)$.
\section{Proofs}
\subsection{Proof of Proposition \ref{prop:ReducedMaxmin}}
For the maximin welfare policy we have
	\begin{align*}
		\min_{(y_0,y_1)\in\mathcal{Y}} \mathbb{E}_{P_X}\left[ \pi(X) \cdot y_{\pi(X)}(X) \right]&=\mathbb{E}_{P_X}\left[ \min_{(y_0(x),y_1(x))\in\mathcal{Y}(x)} \pi(X) \cdot y_{\pi(X)}(X) \right]\\&= \mathbb{E}_{P_X}\left[  \pi(X) \cdot \underline{y}_1(X) +(1-\pi(X))\cdot \underline{y}_0(X)  \right],
	\end{align*}
	where the first equality is justified by \ref{rectangularY}. Thus we have \begin{align*}
	\argmax_{\pi \in \Pi } \min_{(y_0,y_1) \in \mathcal{Y} } W_\tau(\pi)= \argmax_{\pi \in \Pi }\mathbb{E}_{P_X}\left[ \pi(X) \cdot \underline{y}_1(X) +(1-\pi(X))\cdot \underline{y}_0(X)  \right].
\end{align*}
For the maximin impact policy we have
\begin{align*}
		\min_{\tau\in\mathcal{T}} \mathbb{E}_{P_X}\left[ \pi(X) \cdot {\tau}(X) \right]=\mathbb{E}_{P_X}\left[ \min_{\tau\in\mathcal{T}} \pi(X) \cdot {\tau}(X) \right]= \mathbb{E}_{P_X}\left[  \pi(X) \cdot \underline{\tau}(X) \right].
	\end{align*}
where the first equality is justified by Assumption \ref{rectangularTau}. Thus we have 
\begin{align*}
	\argmax_{\pi \in \Pi } \min_{\tau \in \mathcal{T} } W_\tau(\pi)= \argmax_{\pi \in \Pi }\mathbb{E}_{P_X}\left[  \pi(X) \cdot \underline{\tau}(X) \right].
\end{align*}
The statement of the proposition again follows from the invariance of the maximizer to positive affine transformations of the objective function.

\subsection{Proof of Proposition \ref{prop:minimaxRegret}}
We notice that
	\begin{align*}
	&\max_{\tau \in \mathcal{T} }\left( \max_{\pi\, : \, \mathcal{X}\to\{0,1\}}W_\tau(\pi) - W_\tau(\pi)\right)\\ &= \max_{\tau \in \mathcal{T}} \E_{P_X}\left[\left(\frac{1}{2}+\frac{1}{2}\text{sgn}(\tau(X))-\pi(X)\right)\cdot \tau(X)\right] \\
	&= \E_{P_X}\left[\max_{\tau \in \mathcal{T}}\left(\frac{1}{2}+\frac{1}{2}\text{sgn}(\tau(X)) -\pi(X)\right)\cdot \tau(X)\right] \\
	&= \E_{P_X}\bigg[\underbrace{\max_{\tau \in \mathcal{T}}\mathbbm{1}\big\{\underline{\tau}(x) \geq0 \big\}\cdot\left(1-\pi(X)\right)\cdot \tau(X)}_{=(1-\pi)\cdot\mathbbm{1}\left\{\underline{\tau}(x) \geq0 \right\}\cdot\overline{\tau}(X)}\bigg]+\E_{P_X}\bigg[\underbrace{\max_{\tau \in \mathcal{T}}\mathbbm{1}\big\{\overline{\tau}(x) \leq0 \big\}\cdot-\pi(X)\cdot \tau(X)}_{=-\pi\cdot\mathbbm{1}\left\{\underline{\tau}(x) \geq0 \right\}\cdot\underline{\tau}(X)}\bigg]\\
	 &\qquad + \E_{P_X}\bigg[\underbrace{\max_{\tau \in \mathcal{T}}\mathbbm{1}\big\{\underline{\tau}(x)<0<\overline{\tau}(x)  \big\}\cdot\left( \frac{1}{2}+\frac{1}{2}\text{sgn}(\tau(X))-\pi(X)\right) \cdot \tau(X)}_{= -\pi\cdot\mathbbm{1}\left\{\underline{\tau}(X) <0<\overline{\tau}(X) \right\}(\overline{\tau}(X)-\underline{\tau}(X))+\mathbbm{1}\left\{\underline{\tau}(X) <0<\overline{\tau}(X) \right\}\cdot\overline{\tau}(X)}\bigg]\\
	 &=-\E_{P_X}\left[ \pi \cdot \bigg(\mathbbm{1}\big\{\underline{\tau}(X) \geq0 \big\}\cdot\overline{\tau}(X) + \mathbbm{1}\big\{\overline{\tau}(X) \leq0 \big\}\cdot \underline{\tau}(X) + \mathbbm{1}\big\{\underline{\tau}(X)<0<\overline{\tau}(X)\big\}\cdot\big(\overline{\tau}(X) -\underline{\tau}(X) \big) \bigg) \right] \\
	 &\qquad + \E_{P_X}\left[\overline{\tau}(X)\cdot\mathbbm{1}\big\{\underline{\tau}(X) \geq0 \big\} +\mathbbm{1}\left\{\underline{\tau}(X) <0<\overline{\tau}(X) \right\}\cdot\overline{\tau}(X)\right],
	\end{align*}
	where the first equality uses the fact $\argmax_{\pi\, : \, \mathcal{X}\to\{0,1\}}W_\tau(\pi)=\frac{1}{2}+\frac{1}{2}\text{sign}(\tau(X))$, and the second equality uses Assumption \ref{rectangularTau}. The statement of the proposition then follows from the invariance of the maximizer to positive affine transformations of the objective function.

\subsection{Proof of Proposition \ref{prop:regretERMbound}}
We being by decomposing regret as follows:
\begin{align}\label{eq:basicRegretDecompAppendix}
	Q(\pi_n^*) - Q(\widehat{\pi}_n) = {\Big[Q(\pi_n^*) - {Q}^{\texttt{NO}}_n(\pi_n^*) \Big]}+ \Big[Q^{\texttt{NO}}_n(\pi_n^*) - \widehat{Q}^{\texttt{NO}}_n(\widehat{\pi}_n)\Big] +\Big[\widehat{Q}^{\texttt{NO}}_n(\widehat{\pi}_n) - Q(\widehat{\pi}_n) \Big].
\end{align}
The first term is zero in expectation. The second term can be upper bounded as \begin{align*}
	\Big[Q^{\texttt{NO}}_n(\pi_n^*) - \widehat{Q}^{\texttt{NO}}_n(\widehat{\pi}_n)\Big] \leq \Big[Q^{\texttt{NO}}_n(\pi^*_n) - \widehat{Q}^{\texttt{NO}}_n(\pi_n^*)\Big] +\Big[\widehat{Q}^{\texttt{NO}}_n(\pi_n^*) - \widehat{Q}^{\texttt{NO}}_n(\widehat{\pi}_n)\Big] \leq  \sup_{\pi \in \Pi_n}\left| Q^{\texttt{NO}}_n(\pi) - \widehat{Q}^{\texttt{NO}}_n(\pi) \right|,
\end{align*} where we have used that $\widehat{Q}^{\texttt{NO}}_n(\pi^*) - \widehat{Q}^{\texttt{NO}}_n(\widehat{\pi}_n)\leq 0$, which follows from $\widehat{\pi}_n$ being the maximizer of $\widehat{Q}^{\texttt{NO}}_n(\cdot)$. The third term can be further expanded and upper bounded as follows
\begin{equation*}
	\widehat{Q}^{\texttt{NO}}_n(\widehat{\pi}_n) - Q(\widehat{\pi}_n) \leq \sup_{\pi \in \Pi_n}\left|\widehat{Q}^{\texttt{NO}}_n(\pi) - Q^{\texttt{NO}}_n(\pi)\right| +\sup_{\pi \in \Pi_n}\left|Q^{\texttt{NO}}_n(\pi) - Q(\pi)\right| .
\end{equation*}
Using the last two displays and taking expectations in \eqref{eq:basicRegretDecompAppendix} yields the desired conclusion.
\subsection{Proof of Lemma 1}
We will establish each of the following bounds in turn:
\begin{align*}
\mathbb{E}\left[\sup_{\pi\in\Pi_n}|A_{0}(\pi)|\right] &= O\left(\sqrt{{\rm VC}(\Pi_n)\cdot\frac{r_{\kappa_n}}{n^{3/2}}} + \frac{r_{\kappa_n}}{n^{1/2}} \right),\\
\mathbb{E}\left[\sup_{\pi\in\Pi_n}|A_{1,\ell}(\pi)|\right] &= O\left(\sqrt{{\rm VC}(\Pi_n)\cdot\frac{r_{\kappa_n}}{n^{3/2}}} + \frac{r_{\kappa_n}}{n^{1/2}} \right),\\
\mathbb{E}\left[\sup_{\pi\in\Pi_n}|A_{2,\ell}(\pi)|\right] &= O\left(\sqrt{\frac{{\rm VC}(\Pi_n)}{n}} \right)\\
\mathbb{E}\left[\sup_{\pi\in\Pi_n}|A_{3,\ell}(\pi)|\right] &= O\left(\left(\frac{r_{\kappa_n}}{n}\right)^{\frac{\gamma+1}{\gamma+2}} \right)
\end{align*}
Combining the above through decomposition \eqref{eq:DecompositionQhat-Qtilde} gives the desired final bound.
\subsubsection*{Bound for $A_0$ and $A_{1,\ell}$}
We prove a bound for $\sup_{\pi_\in \Pi_n}|A_{1,\ell}(\pi)|$; it will be immediate that $A_0(\pi)$ obeys the same bound. We begin with the following decomposition
\begin{align*}
A_{1,\ell}(\pi)&=\frac{1}{n}\sum_{i=1}^n(2\pi(X_i)-1)\cdot\langle  \widehat \alpha_i - \alpha_i, U_i -g(V_i)\rangle,\, &\left(=B_1(\pi)\right),\\&+ \frac{1}{n}\sum_{i=1}^n(2\pi(X_i)-1)\cdot \langle \widehat \alpha_i-\alpha_i, g_i -\widehat g_i \rangle, \, &(=B_2(\pi)),\\&+ \frac{1}{n}\sum_{i=1}^n(2\pi(X_i)-1)\cdot\left(\varphi_\ell(\widehat g^{-k(i)};X_i) - \varphi_\ell(g;X_i) + \langle \alpha_i, g_i - \widehat g_i \rangle\right),\, &(=B_3(\pi)),
\end{align*}
where we have used the shorthand notation $\widehat g_i:= \widehat g^{-k(i)}(V_i) $, $g_i=g(V_i)$, $\widehat \alpha_i := (\{\widehat g^{-k(i)},f^{-k(i)}\};V_i)$, $\alpha_i := (\{ g,f\};V_i)$.
Starting with $B_1(\pi)$, the contribution of the $k$-th fold is 
\begin{align*}
B_1^{(k)}(\pi)=\frac{1}{n}\sum_{i:k(i)=k} (2\pi(X_i)-1)\cdot \langle \widehat \alpha_i - \alpha_i , \varepsilon_i\rangle \cdot\mathbbm{1}\{\widehat \varphi_i\geq0\}.
\end{align*}
The sample-splitting procedure guarantees that $\{\widehat{g}^{-k(i)},\widehat{f}^{-k(i)}\}$ only depend on data from the remaining $K-1$ folds, and thus conditioning on these estimates for the nuisance components makes $ B_1^{(k)}(\pi)$ a sum of independent mean-zero terms, in light of \begin{align*}\mathbb{E}\left[ U_i -  g(V_i) \mid V_i,\widehat{g}^{-k(i)},\widehat{f}^{-k(i)} \right]=0.\end{align*} Furthermore, the terms are also sub-Gaussian since it is a linear combination of sub-Gaussian random variables with bounded weights w.p.a 1, in light of
\begin{align*}
\norm{\mathbbm{1}\{\widehat \varphi_i\geq0\}\cdot(\widehat \alpha _i - \alpha_i) }_{L_\infty(P_V)} \leq \norm{\widehat g^{-k}(V) -g(V)}_{L_\infty(P_V)} + \norm{\widehat f^{-k}(V) -f(V)}_{L_\infty(P_V)}\leq 2\cdot \mathcal{C}_2\cdot\mathcal{C}_3,
\end{align*}
w.p.a 1, where the first inequality uses Assumption \ref{ass:Regularity}(i) and the second inequality uses Assumption \ref{ass:Regularity}(iii). Having computed the variance of $B_1^{(k)}(\pi)$ conditional on $\left(\widehat g^{-k},\widehat f^{-k}\right)$
\begin{align*}
	V_n(k)=\mathbb{E}\left[(\widehat{\alpha}^{-k}_i-\alpha_i)'\,\Sigma(V_i)\,(\widehat{\alpha}^{-k}_i-\alpha_i)\cdot\mathbbm{1}\left\{\widehat{\varphi}_i \geq0\right\} \mid \widehat{g}^{-k},\widehat{f}^{-k}\right],
\end{align*}
we can apply Corollary 3 in \cite{atheywager2021} to establish the bound
\begin{align}\label{boundATHEYWAGERproof}
	\frac{n}{n_k}\E\left[\sup_{\pi \in \Pi}\left|B_1^{(k)}(\pi)\right|\mid \hat{g}^{-k}\right]= O\left(\sqrt{V_n(k)\frac{\text{VC}(\Pi_n)}{n_k}} \right),
\end{align}
where $n_k$ denotes the number of observations in the $k$-th fold. Using Assumptions \ref{ass:Regularity}(ii) and \ref{ass:L2rates}, we have
\begin{equation}\label{boundVk}
\begin{aligned}
	\E\left[V_n(k)\right] &\leq \E_{P_n}\left[\overline \lambda \cdot \norm{\widehat \alpha_i -\alpha_i}^2_{L_2(P_V)} \right]\\
	&\leq 2\cdot\overline{\lambda}\cdot \mathcal{C}_{2,\alpha}^2\cdot\mathbb{E}_{P_n}\left[\norm{\widehat g^{-k(i)} - g }_{L_2(P_V)}^2 + \norm{\widehat f^{-k(i)} - f }_{L_2(P_V)}^2 \right] \\	
	&=O\left(\frac{r_{\kappa_n}}{\sqrt{n}} \right).  
\end{aligned}
\end{equation}
Finally, we apply (\ref{boundATHEYWAGERproof}) repeatedly for each of the $K$ data-folds and using Jensen's Inequality and (\ref{boundVk}) and obtain the final bound
\begin{align}\label{boundB1}
	\E\left[\sup_{\pi \in \Pi} \left|B_1(\pi)\right|\right]=O\left(\sqrt{\text{VC}(\Pi_n)\cdot\frac{r_{\kappa_n}}{n^{3/2}}} \right).
\end{align}
We now turn to $B_2(\pi)$, for which we have
\begin{align*}
B_2(\pi)&= \frac{1}{n}\sum_{i=1}^n(2\pi(X_i)-1)\cdot \langle \widehat \alpha_i - \alpha_i , \widehat g_i - g_i\rangle \cdot\mathbbm{1}\{\widehat \varphi_i\geq0\}\\
&= \sum_{j=1}^J\left[\frac{1}{n}\sum_{i=1}^n(2\pi(X_i)-1)\cdot ( \widehat \alpha^{(j)}_i - \alpha^{(j)}_i) \cdot(\widehat g_i^{(j)} - g_i) \cdot\mathbbm{1}\{\widehat \varphi_i\geq0\}\right]\\
&\leq \sum_{j=1}^J\left[\frac{1}{n}\sum_{i=1}^n |\widehat \alpha^{(j)}_i - \alpha^{(j)}_i| \cdot|\widehat g_i^{(j)} - g_i|\right]\\
&\leq \sum_{j=1}^J \sqrt{\frac{1}{n}\sum_{i=1}^n \left(\widehat \alpha^{(j)}_i - \alpha^{(j)}_i\right)^2}\times \sqrt{\frac{1}{n}\sum_{i=1}^n \left(g_i^{(j)} - g_i\right)^2},
\end{align*}
where the last inequality uses Cauchy-Schwarz inequality. This bound does not depend on $\pi$ and thus holds uniformly over $\Pi_n$. We then apply Cauchy-Schwarz again and use Asuumption \ref{ass:L2rates} to verify that
\begin{align*}
	\E\left[\sup_{\pi \in \Pi} \left|B_2(\pi)\right|\right]&\leq \sum_{j=1}^J \mathbb{E}_{P_n}\left[\norm{\widehat \alpha^{(j)}_i - \alpha^{(j)}_i}^2_{L_2(P_V)}\right]^{1/2} \times \mathbb{E}_{P_n}\left[\norm{\widehat g^{(j)}_i - g^{(j)}_i}^2_{L_2(P_V)}\right]^{1/2} ,\\	
	&\leq J\cdot \mathbb{E}_{P_n}\left[\norm{\widehat \alpha_i - \alpha_i}^2_{L_2(P_V)}\right]^{1/2} \times\mathbb{E}_{P_n}\left[\norm{\widehat g_i - g_i}^2_{L_2(P_V)}\right]^{1/2}  \\
	&\lesssim	 J\cdot\mathbb{E}_{P_n}\left[\norm{\widehat f^{-k(i)} - f}^2_{L_2(P_V)}+\norm{\widehat g^{-k(i)} - g}^2_{L_2(P_V)}\right]^{1/2} \times\mathbb{E}_{P_n}\left[\norm{\widehat g^{-k(i)} - g}^2_{L_2(P_V)}\right]^{1/2} \\
	&=O\left(\frac{r_{\kappa_n}}{\sqrt{n}} \right).
\end{align*}
We now turn to $B_3(\pi)$. We begin by considering the following telescoping
\begin{align*}
\widehat g_i - g_i = \sum_{j=1}^J \left[(g^{(\bullet:j-1)}_i ,\widehat g^{(j:\bullet)}_i) - (g^{(\bullet:j)}_i ,\widehat g^{(j+1:\bullet)}_i)\right]= \sum_{j=1}^J\left(0,0,\dots,\widehat g^{(j)}_i-g^{(j)}_i,0,\dots,0\right),
\end{align*}
where $g^{(\bullet:j)}$ and $g^{(j:\bullet)}$  denote, respectively, the first and last $j$ entries of $g(V_i)$, where we adopt the convention $g^{(\bullet:0)}=g^{(J+1:\bullet)}=\emptyset$. We can therefore decompose $B_3(\pi)$ as follows:
\begin{align*}
&B_3(\pi)=\sum_{j=1}^J \frac{1}{n}\sum_{i=1}^n(2\pi(X_i)-1)\\ &\times \left(\varphi(( g^{(\bullet:j-1)}, \widehat g^{(j:\bullet)}_i);X_i) - \varphi((g^{(\bullet:j)}_i ,\widehat g^{(j+1:\bullet)}_i);X_i) - \alpha^{(j)}(\{(g^{(\bullet:j-1)},\widehat g^{(j:\bullet}),f\})\cdot \left(\widehat g^{(j)}_i- g^{(j)}_i \right)\right)\cdot\mathbbm{1}\{\widehat \varphi_i\geq0\}\\
&- \sum_{j=1}^J\frac{1}{n}\sum_{i=1}^n(2\pi(X_i)-1)\cdot\left[\left( \alpha^{(j)}_i-\alpha^{(j)}(\{(g^{(\bullet:j-1},\widehat g^{(j:\bullet)}),f\})\right)\cdot \left(\widehat{g}^{(j)}_i-g^{(j)}_i\right)\right]\cdot\mathbbm{1}\{\widehat \varphi_i\geq0\}.
\end{align*}
By the definition of the Riesz-representer and cross-fitting we have
\begin{align*}\mathbb{E}\Big[\varphi(( g^{(\bullet:j-1)}, &\widehat g^{(j:\bullet)}_i);X_i) - \varphi((g^{(\bullet:j)}_i ,\widehat g^{(j+1:\bullet)}_i);X_i)\\ &- \alpha^{(j)}(\{(g^{(\bullet:j-1)},\widehat g^{(j:\bullet}),f\},W_i)\cdot \left(\widehat g^{(j)}_i- g^{(j)}_i \right)\mid V_i,\widehat{g}^{-k(i)},\widehat{f}^{-k(i)} \Big]=0,
\end{align*}
where we have used the property $\alpha^{(j)}_\ell(\{(\widetilde g_{-j},\widetilde g_{j}),\widetilde f\}, x)=\alpha^{(j)}_\ell(\{\widetilde g_{-j},\widetilde f\}, x)$. Furthermore, the term within the expectation operator is sub-Gaussian since uniformly bounded by Assumption \ref{ass:Regularity}(iii). Therefore the first term in the expansion of $B_3(\pi)$ can be controlled uniformly using identical arguments as for $B_1(\pi)$ and obeys the same bound. The second term in the expansion of $B_3(\pi)$ can be bounded with identical arguments as for $B_2(\pi)$ and obeys the same bound. We therefore conclude that
\begin{align*}
	\E\left[\sup_{\pi \in \Pi} \left|B_3(\pi)\right|\right]=O\left(\sqrt{\text{VC}(\Pi_n)\cdot\frac{r_{\kappa_n}}{n^{3/2}}}+ \frac{r_{\kappa_n}}{\sqrt{n}}\right).
\end{align*}
Combining the bounds for $B_1(\pi)$,$B_2(\pi)$ and $B_3(\pi)$ via the triangle inequality finally gives the desired bound for $\mathbb{E}\left[\sup_{\pi\in\Pi_n}|A_{1,\ell}(\pi)|\right]$.
\subsubsection*{Bound for  $A_{2,\ell}$}
We first notice that \begin{align}\mathbb{E}\left[\phi_{\ell}(\{g,f\};W_i) \cdot\left( \mathbbm{1}\left\{\varphi_\ell(\widehat g^{-k(i)};X_i) \geq0\right\} - \mathbbm{1}\left\{\varphi_\ell(g;X_i) \geq0  \right\} \right)\mid V_i, \widehat g^{-k(i)} \right]=0,\end{align}
by the mean-zero property of the influence function adjustments $\phi_\ell$ and cross-fitting. Furthermore, the term inside expectation is sub-Gaussian by uniform boundedness of the Riesz-representer, guaranteed by Assumption \ref{ass:Regularity}(iii). Thus we can use similar arguments to those used for $B_1(\pi)$ to show
\begin{align*}
\mathbb{E}\left[\sup_{\pi\in\Pi_n}|A_{2,\ell}(\pi)|\right]=O\left(\sqrt{\frac{{\rm VC}(\Pi_n)}{n}} \right).
\end{align*}
\subsubsection*{Bound for  $A_{3,\ell}$}
We begin by noticing that $\varphi_\ell(g;X_i) \left( \mathbbm{1}\left\{\varphi_\ell(\widehat g^{-k(i)};X_i) \geq0\right\} - \mathbbm{1}\left\{\varphi_\ell(g;X_i) \geq0  \right\} \right)\leq 0$ and thus, since the ``never treat" policy belongs to any policy class $\Pi$ for which $\text{VC}(\Pi)\geq 1$, we have
\begin{align*}
	\sup_{\pi\in\Pi_n}A_{3,\ell}(\pi)=\frac{1}{2n}\sum_{i=1}^n \left|\varphi_\ell(g;X_i) \cdot \left( \mathbbm{1}\left\{\varphi_\ell(\widehat g^{-k(i)};X_i) \geq0\right\} - \mathbbm{1}\left\{\varphi_\ell(g;X_i) \geq0  \right\} \right)\right|,
\end{align*}
and thus we obtain the uniform bound\footnote{For a policy class of zero VC-dimension, (\ref{C2firstStep})  holds as an inequality.} 
\begin{align}\label{C2firstStep}
	\E\left[\sup_{\pi\in\Pi_n}A_{3,\ell}(\pi)\right]=\frac{1}{2} \E\Bigg[\left|\varphi_\ell(g;X_i) \cdot \left( \mathbbm{1}\left\{\varphi_\ell(\widehat g^{-k(i)};X_i) \geq0\right\} - \mathbbm{1}\left\{\varphi_\ell(g;X_i) \geq0  \right\} \right)\right|\Bigg].
\end{align}
For the RHS in (\ref{C2firstStep}), we closely follow Lemma 5.2 in \citet{audibertTsybakov2007}, but we report the steps of the proof for completeness. For $\gamma>0$ and any $t>0$ we have
\begin{align*}
	&\E\Bigg[\left|\varphi_\ell(g;X_i) \left( \mathbbm{1}\left\{\varphi_\ell(\widehat g^{-k(i)};X_i) \geq0\right\} - \mathbbm{1}\left\{\varphi_\ell(g;X_i) \geq0  \right\} \right)\right|\Bigg]\\
	&\leq \E\Bigg[\left|\varphi_\ell(g;X_i)\right| \cdot \mathbbm{1}\left\{\left|\varphi_\ell(\widehat g^{-k(i)};X_i)-\varphi_\ell(g;X_i) \right| \geq\left|\varphi_\ell(g;X_i)\right|\right\}\Bigg]\\
	&\leq\E\Bigg[\left|\varphi_\ell(g;X_i)\right| \cdot \mathbbm{1}\left\{\left|\varphi_\ell(\widehat g^{-k(i)};X_i)-\varphi_\ell(g;X_i) \right| \geq\left|\varphi_\ell(g;X_i)\right|\right\}\cdot\mathbbm{1}\left\{0<\left|\varphi_\ell(g;X_i)\right| \leq t \right\}\Bigg]\\ & \qquad  + \E\Bigg[\left|\varphi_\ell(g;X_i)\right| \cdot \mathbbm{1}\left\{\left|\varphi_\ell(\widehat g^{-k(i)};X_i)-\varphi_\ell(g;X_i) \right| \geq\left|\varphi_\ell(g;X_i)\right|\right\}\cdot\mathbbm{1}\left\{\left|\varphi_\ell(g;X_i)\right| > t \right\}\Bigg]\\	&\leq\E\Bigg[\left|\varphi_\ell(\widehat g^{-k(i)};X_i)-\varphi_\ell(g;X_i)\right| \cdot\mathbbm{1}\left\{0<\left|\varphi_\ell(g;X_i)\right| \leq t  \right\}\Bigg]\\ & \qquad + \E\Bigg[\left|\varphi_\ell(\widehat g^{-k(i)};X_i)-\varphi_\ell(g;X_i)\right| \cdot \mathbbm{1}\left\{\left|\varphi_\ell(\widehat g^{-k(i)};X_i)-\varphi_\ell(g;X_i) \right| > t\right\}\Bigg]\\
	&\leq \E\left[\left( \varphi_\ell(\widehat g^{-k(i)};X_i)-\varphi_\ell(g;X_i)\right) ^2\right]^{1/2} \cdot \mathbb{P}\big(0< \left|\varphi_\ell(g;X_i)\right|\leq t  \big)^{1/2} + \frac{\E\left[\left( \varphi_\ell(\widehat g^{-k(i)};X_i)-\varphi_\ell(g;X_i)\right) ^2\right]}{t}\\
	&\leq C_0^{1/2} \E\left[\left( \varphi_\ell(\widehat g^{-k(i)};X_i)-\varphi_\ell(g;X_i)\right) ^2\right]^{1/2}t^{\gamma/2} + \frac{\E\left[\left( \varphi_\ell(\widehat g^{-k(i)};X_i)-\varphi_\ell(g;X_i)\right) ^2\right]}{t},
\end{align*}
where the penultimate inequality uses Cauchy-Schwarz and Markov inequalities, and the last inequality uses the Margin Assumption. Minimizing the last display over $t$ gives
\begin{align*}
		\E\left[\sup_{\pi\in\Pi_n}A_{3,\ell}(\pi)\right]&\leq (\gamma+2)\cdot\left( \frac{2}{\gamma}\right)^{\gamma/(\gamma+2)}\cdot \mathcal{C}_m^{1/(\gamma+2)}\cdot\E\left[\left( \varphi_\ell(\widehat g^{-k(i)};X_i)-\varphi_\ell(g;X_i)\right) ^2\right]^{\frac{\gamma+1}{\gamma+2}}\\
			&\leq  (\gamma+2)\cdot\left( \frac{2}{\gamma}\right)^{\gamma/(\gamma+2)}\cdot \mathcal{C}_m^{1/(\gamma+2)}\cdot\mathcal{C}_{2,\varphi}^{\frac{2(\gamma+1)}{\gamma+2}}\cdot\E_{P_n}\left[\norm{\widehat g^{-k} -g}_{L_2(P_X)}^2\right]^{\frac{\gamma+1}{\gamma+2}}
\end{align*} 
For $\gamma=0$, a similar argument gives
\begin{align*}
	&\E\Bigg[\left|\varphi_\ell(g;X_i) \left( \mathbbm{1}\left\{\varphi_\ell(\widehat g^{-k(i)};X_i) \geq0\right\} - \mathbbm{1}\left\{\varphi_\ell(g;X_i) \geq0  \right\} \right)\right|\Bigg]\\
	&\leq\E\Bigg[\left|\varphi_\ell(\widehat g^{-k(i)};X_i)-\varphi_\ell(g;X_i)\right| \cdot\mathbbm{1}\left\{0<\left|\varphi_\ell(g;X_i)\right| \leq t  \right\}\Bigg]\\ & \qquad \qquad \qquad + \E\Bigg[\left|\varphi_\ell(\widehat g^{-k(i)};X_i)-\varphi_\ell(g;X_i)\right| \cdot \mathbbm{1}\left\{\left|\varphi_\ell(\widehat g^{-k(i)};X_i)-\varphi_\ell(g;X_i) \right| > t\right\}\Bigg]\\
	& \leq 2 \mathbb{E}_{P_n}\left[\norm{\varphi_\ell(\widehat g^{-k};X)-\varphi_\ell(g;X)}_{L_2(P_X)}^2\right]^{1/2}\\
	&\leq 2 \cdot \mathcal{C}_{2,\varphi}^2\cdot\mathbb{E}_{P_n}\left[\norm{\widehat g^{-k}-g}_{L_2(P_X)}^2\right]^{1/2}.
\end{align*}
Combining the cases $\gamma>0$ and $\gamma=0$, and using the $L_2$-risk bounds for $\widehat g^{-k}$ from Assumption \ref{ass:L2rates} we finally get
\begin{align*}
	\E\left[\sup_{\pi\in\Pi_n}A_{3,\ell}(\pi)\right]=O\left(\left(\frac{ r_{\kappa_n}}{\sqrt{n}}\right)^{\frac{\gamma+1}{\gamma+2}} \right).
\end{align*}

\subsection{Proof of Theorem \ref{mainTheorem}}
The Neyman-orthogonalized score $\Gamma^{\texttt{NO}}(\{g,f\};W_i)$ satisfies the assumptions of Corollary 3 in \cite{atheywager2021}, and thus it can be applied verbatim to show that
\begin{align*}
\mathbb{E}\left[\sup_{\pi \in \Pi_n}\left|Q^{\texttt{NO}}_n(\pi) - Q(\pi)\right|\right] =O\left(\sqrt{\frac{{\rm VC}(\Pi_n)}{n}}\right).\end{align*}
Combining the above bound with Lemma \ref{mainLemma} via Proposition \ref{prop:regretERMbound} gives the statement of the theorem.
\end{appendices}
~\\

\newpage
\setlength{\bibsep}{2pt}
\bibliographystyle{chicago}
\bibliography{refs}

\begin{thebibliography}{}

\bibitem[\protect\citeauthoryear{Abadie, Angrist, and Imbens}{Abadie
  et~al.}{2002}]{abadieAngrist2002}
Abadie, A., J.~Angrist, and G.~Imbens (2002).
\newblock Instrumental variables estimates of the effect of subsidized training
  on the quantiles of trainee earnings.
\newblock {\em Econometrica\/}~{\em 70\/}(1), 91--117.

\bibitem[\protect\citeauthoryear{Adjaho and Christensen}{Adjaho and
  Christensen}{2022}]{christensenAdjaho}
Adjaho, C. and T.~Christensen (2022).
\newblock Externally valid treatment choice.

\bibitem[\protect\citeauthoryear{Arlot and Bartlett}{Arlot and
  Bartlett}{2011}]{arlotBartlett2011}
Arlot, S. and P.~L. Bartlett (2011).
\newblock {Margin-adaptive model selection in statistical learning}.
\newblock {\em Bernoulli\/}~{\em 17\/}(2), 687 -- 713.

\bibitem[\protect\citeauthoryear{Athey and Wager}{Athey and
  Wager}{2021}]{atheywager2021}
Athey, S. and S.~Wager (2021).
\newblock Policy learning with observational data.
\newblock {\em Econometrica\/}~{\em 89\/}(1), 133--161.

\bibitem[\protect\citeauthoryear{Audibert and Tsybakov}{Audibert and
  Tsybakov}{2007}]{audibertTsybakov2007}
Audibert, J.-Y. and A.~B. Tsybakov (2007).
\newblock {Fast learning rates for plug-in classifiers}.
\newblock {\em The Annals of Statistics\/}~{\em 35\/}(2), 608 -- 633.

\bibitem[\protect\citeauthoryear{Balke and Pearl}{Balke and
  Pearl}{1997}]{balkePearl2007}
Balke, A. and J.~Pearl (1997).
\newblock Bounds on treatment effects from studies with imperfect compliance.
\newblock {\em Journal of the American Statistical Association\/}~{\em
  92\/}(439), 1171--1176.

\bibitem[\protect\citeauthoryear{Berger}{Berger}{1985}]{Berger}
Berger, J.~O. (1985).
\newblock {\em Statistical Decision Theory and Bayesian Analysis}.
\newblock Springer Series in Statistics, 2nd edition.

\bibitem[\protect\citeauthoryear{Bloom, Orr, Bell, Cave, Doolittle, Lin, and
  Bos}{Bloom et~al.}{1997}]{bloom1997}
Bloom, H.~S., L.~L. Orr, S.~H. Bell, G.~Cave, F.~Doolittle, W.~Lin, and J.~M.
  Bos (1997).
\newblock The benefits and costs of jtpa title ii-a programs: Key findings from
  the national job training partnership act study.
\newblock {\em The Journal of Human Resources\/}~{\em 32\/}(3), 549--576.

\bibitem[\protect\citeauthoryear{Byambadalai}{Byambadalai}{2022}]{Byambadalai2022}
Byambadalai, U. (2022).
\newblock Identification and inference for welfare gains without
  unconfoundedness.

\bibitem[\protect\citeauthoryear{Chamberlain}{Chamberlain}{2011}]{chamberlain2011}
Chamberlain, G. (2011, 09).
\newblock {1011 Bayesian Aspects of Treatment Choice}.
\newblock In {\em {The Oxford Handbook of Bayesian Econometrics}}. Oxford
  University Press.

\bibitem[\protect\citeauthoryear{Chen}{Chen}{2007}]{chen2007}
Chen, X. (2007).
\newblock Large sample sieve estimation of semi-nonparametric models.
\newblock Volume~6 of {\em Handbook of Econometrics}, pp.\  5549--5632.
  Elsevier.

\bibitem[\protect\citeauthoryear{Chernozhukov, Escanciano, Ichimura, Newey, and
  Robins}{Chernozhukov et~al.}{2022}]{chernozhukov2018}
Chernozhukov, V., J.~C. Escanciano, H.~Ichimura, W.~K. Newey, and J.~M. Robins
  (2022).
\newblock Locally robust semiparametric estimation.
\newblock {\em Econometrica\/}~{\em 90\/}(4), 1501--1535.

\bibitem[\protect\citeauthoryear{Chernozhukov, Lee, and Rosen}{Chernozhukov
  et~al.}{2013}]{chernozhukovIntersectionBounds}
Chernozhukov, V., S.~Lee, and A.~M. Rosen (2013).
\newblock Intersection bounds: Estimation and inference.
\newblock {\em Econometrica\/}~{\em 81\/}(2), 667--737.

\bibitem[\protect\citeauthoryear{Christensen, Moon, and
  Schorfheide}{Christensen et~al.}{2022}]{christensenMoonSchor2020}
Christensen, T., H.~R. Moon, and F.~Schorfheide (2022).
\newblock Optimal discrete decisions when payoffs are partially identified.

\bibitem[\protect\citeauthoryear{Cui and Tchetgen}{Cui and
  Tchetgen}{2021}]{cuiTchetgen2021}
Cui, Y. and E.~T. Tchetgen (2021).
\newblock A semiparametric instrumental variable approach to optimal treatment
  regimes under endogeneity.
\newblock {\em Journal of the American Statistical Association\/}~{\em
  116\/}(533), 162--173.
\newblock PMID: 33994604.

\bibitem[\protect\citeauthoryear{Dehejia}{Dehejia}{2005}]{dehejia2005}
Dehejia, R.~H. (2005).
\newblock Program evaluation as a decision problem.
\newblock {\em Journal of Econometrics\/}~{\em 125\/}(1), 141--173.
\newblock Experimental and non-experimental evaluation of economic policy and
  models.

\bibitem[\protect\citeauthoryear{Fang and Santos}{Fang and
  Santos}{2018}]{fangSantos2019}
Fang, Z. and A.~Santos (2018, 09).
\newblock {Inference on Directionally Differentiable Functions}.
\newblock {\em The Review of Economic Studies\/}~{\em 86\/}(1), 377--412.

\bibitem[\protect\citeauthoryear{Farrell}{Farrell}{2015}]{farrell2015}
Farrell, M.~H. (2015).
\newblock Robust inference on average treatment effects with possibly more
  covariates than observations.
\newblock {\em Journal of Econometrics\/}~{\em 189\/}(1), 1--23.

\bibitem[\protect\citeauthoryear{Farrell, Liang, and Misra}{Farrell
  et~al.}{2021}]{farrell2020}
Farrell, M.~H., T.~Liang, and S.~Misra (2021).
\newblock Deep neural networks for estimation and inference.
\newblock {\em Econometrica\/}~{\em 89\/}(1), 181--213.

\bibitem[\protect\citeauthoryear{Foster and Syrgkanis}{Foster and
  Syrgkanis}{2019}]{fosterSyrgkanis}
Foster, D.~J. and V.~Syrgkanis (2019).
\newblock Orthogonal statistical learning.

\bibitem[\protect\citeauthoryear{Han}{Han}{2019}]{han2019}
Han, S. (2019).
\newblock Optimal dynamic treatment regimes and partial welfare ordering.

\bibitem[\protect\citeauthoryear{Heckman and Vytlacil}{Heckman and
  Vytlacil}{2001}]{heckman2001instrumental}
Heckman, J.~J. and E.~J. Vytlacil (2001).
\newblock Instrumental variables, selection models, and tight bounds on the
  average treatment effect.
\newblock In {\em Econometric Evaluation of Labour Market Policies}, pp.\
  1--15. Springer.

\bibitem[\protect\citeauthoryear{Hirano and Porter}{Hirano and
  Porter}{2009}]{hiranoPorter2009ECMA}
Hirano, K. and J.~R. Porter (2009).
\newblock Asymptotics for statistical treatment rules.
\newblock {\em Econometrica\/}~{\em 77\/}(5), 1683--1701.

\bibitem[\protect\citeauthoryear{Hirano and Porter}{Hirano and
  Porter}{2012}]{hiranoPorter2012ECMA}
Hirano, K. and J.~R. Porter (2012).
\newblock Impossibility results for nondifferentiable functionals.
\newblock {\em Econometrica\/}~{\em 80\/}(4), 1769--1790.

\bibitem[\protect\citeauthoryear{Hirano and Porter}{Hirano and
  Porter}{2020}]{hiranoPorter2020HoE}
Hirano, K. and J.~R. Porter (2020).
\newblock Asymptotic analysis of statistical decision rules in econometrics.
\newblock In S.~N. Durlauf, L.~P. Hansen, J.~J. Heckman, and R.~L. Matzkin
  (Eds.), {\em Handbook of Econometrics, Volume 7A}, Volume~7 of {\em Handbook
  of Econometrics}, pp.\  283--354. Elsevier.

\bibitem[\protect\citeauthoryear{Hurwicz}{Hurwicz}{1951}]{hurwicz}
Hurwicz, L. (1951).
\newblock The generalised bayes-minimax principle: A criterion for
  decision-making under uncertainty.

\bibitem[\protect\citeauthoryear{Ichimura and Newey}{Ichimura and
  Newey}{2022}]{ichimura2015}
Ichimura, H. and W.~K. Newey (2022).
\newblock The influence function of semiparametric estimators.
\newblock {\em Quantitative Economics\/}~{\em 13\/}(1), 29--61.

\bibitem[\protect\citeauthoryear{Imbens and Angrist}{Imbens and
  Angrist}{1994}]{imbensangrist1994}
Imbens, G.~W. and J.~D. Angrist (1994).
\newblock Identification and estimation of local average treatment effects.
\newblock {\em Econometrica\/}~{\em 62\/}(2), 467--475.

\bibitem[\protect\citeauthoryear{Ishihara and Kitagawa}{Ishihara and
  Kitagawa}{2021}]{ishiharaKitagawa}
Ishihara, T. and T.~Kitagawa (2021).
\newblock Evidence aggregation for treatment choice.

\bibitem[\protect\citeauthoryear{Kallus and Zhou}{Kallus and
  Zhou}{2018}]{kallus2018}
Kallus, N. and A.~Zhou (2018).
\newblock Confounding-robust policy improvement.

\bibitem[\protect\citeauthoryear{Kasy}{Kasy}{2016}]{kasy2016}
Kasy, M. (2016, 03).
\newblock {Partial Identification, Distributional Preferences, and the Welfare
  Ranking of Policies}.
\newblock {\em The Review of Economics and Statistics\/}~{\em 98\/}(1),
  111--131.

\bibitem[\protect\citeauthoryear{Kennedy}{Kennedy}{2022}]{kennedy2022}
Kennedy, E.~H. (2022).
\newblock Semiparametric doubly robust targeted double machine learning: a
  review.

\bibitem[\protect\citeauthoryear{Kido}{Kido}{2022}]{kido2022}
Kido, D. (2022).
\newblock Distributionally robust policy learning with wasserstein distance.

\bibitem[\protect\citeauthoryear{Kim, Kwon, Kwon, and Lee}{Kim
  et~al.}{2018}]{simonLeeBounds2018}
Kim, W., K.~Kwon, S.~Kwon, and S.~Lee (2018).
\newblock The identification power of smoothness assumptions in models with
  counterfactual outcomes.
\newblock {\em Quantitative Economics\/}~{\em 9\/}(2), 617--642.

\bibitem[\protect\citeauthoryear{Kitagawa, Lee, and Qiu}{Kitagawa
  et~al.}{2022}]{kitagawaLeeQiu2022}
Kitagawa, T., S.~Lee, and C.~Qiu (2022).
\newblock Treatment choice with nonlinear regret.

\bibitem[\protect\citeauthoryear{Kitagawa, {Montiel Olea}, Payne, and
  Velez}{Kitagawa et~al.}{2020}]{kitagawaEtAl2020}
Kitagawa, T., J.~L. {Montiel Olea}, J.~Payne, and A.~Velez (2020).
\newblock Posterior distribution of nondifferentiable functions.
\newblock {\em Journal of Econometrics\/}~{\em 217\/}(1), 161--175.

\bibitem[\protect\citeauthoryear{Kitagawa, Sakaguchi, and Tetenov}{Kitagawa
  et~al.}{2021}]{kitagawaSakaguchi2021}
Kitagawa, T., S.~Sakaguchi, and A.~Tetenov (2021).
\newblock Constrained classification and policy learning.

\bibitem[\protect\citeauthoryear{Kitagawa and Tetenov}{Kitagawa and
  Tetenov}{2018}]{kitagawatetenov2018}
Kitagawa, T. and A.~Tetenov (2018).
\newblock Who should be treated? empirical welfare maximization methods for
  treatment choice.
\newblock {\em Econometrica\/}~{\em 86\/}(2), 591--616.

\bibitem[\protect\citeauthoryear{Lee}{Lee}{2009}]{lee2009}
Lee, D.~S. (2009, 07).
\newblock {Training, Wages, and Sample Selection: Estimating Sharp Bounds on
  Treatment Effects}.
\newblock {\em The Review of Economic Studies\/}~{\em 76\/}(3), 1071--1102.

\bibitem[\protect\citeauthoryear{Mammen and Tsybakov}{Mammen and
  Tsybakov}{1999}]{mammenTsybakov1999}
Mammen, E. and A.~B. Tsybakov (1999).
\newblock {Smooth discrimination analysis}.
\newblock {\em The Annals of Statistics\/}~{\em 27\/}(6), 1808 -- 1829.

\bibitem[\protect\citeauthoryear{Manski}{Manski}{1990}]{manski1990}
Manski, C.~F. (1990).
\newblock Nonparametric bounds on treatment effects.
\newblock {\em The American Economic Review\/}~{\em 80\/}(2), 319--323.

\bibitem[\protect\citeauthoryear{Manski}{Manski}{2004}]{manski2004}
Manski, C.~F. (2004).
\newblock Statistical treatment rules for heterogeneous populations.
\newblock {\em Econometrica\/}~{\em 72\/}(4), 1221--1246.

\bibitem[\protect\citeauthoryear{Manski}{Manski}{2009}]{manski2009}
Manski, C.~F. (2009).
\newblock Diversified treatment under ambiguity.
\newblock {\em International Economic Review\/}~{\em 50\/}(4), 1013--1041.

\bibitem[\protect\citeauthoryear{Manski}{Manski}{2010}]{manski2010}
Manski, C.~F. (2010).
\newblock Vaccination with partial knowledge of external effectiveness.
\newblock {\em Proceedings of the National Academy of Sciences\/}~{\em
  107\/}(9), 3953--3960.

\bibitem[\protect\citeauthoryear{Manski}{Manski}{2011}]{manski2011}
Manski, C.~F. (2011).
\newblock Choosing treatment policies under ambiguity.
\newblock {\em Annual Review of Economics\/}~{\em 3\/}(1), 25--49.

\bibitem[\protect\citeauthoryear{Manski and Pepper}{Manski and
  Pepper}{2000}]{manskiPepper}
Manski, C.~F. and J.~V. Pepper (2000).
\newblock Monotone instrumental variables: With an application to the returns
  to schooling.
\newblock {\em Econometrica\/}~{\em 68\/}(4), 997--1010.

\bibitem[\protect\citeauthoryear{Mbakop and Tabord-Meehan}{Mbakop and
  Tabord-Meehan}{2021}]{mbakopTabord}
Mbakop, E. and M.~Tabord-Meehan (2021).
\newblock Model selection for treatment choice: Penalized welfare maximization.
\newblock {\em Econometrica\/}~{\em 89\/}(2), 825--848.

\bibitem[\protect\citeauthoryear{Ponomarev}{Ponomarev}{2022}]{ponomarev}
Ponomarev, K. (2022).
\newblock Efficient estimation of directionally differentiable functionals.

\bibitem[\protect\citeauthoryear{Pu and Zhang}{Pu and Zhang}{2021}]{Pu_2021}
Pu, H. and B.~Zhang (2021, mar).
\newblock Estimating optimal treatment rules with an instrumental variable: A
  partial identification learning approach.
\newblock {\em Journal of the Royal Statistical Society: Series B (Statistical
  Methodology)\/}~{\em 83\/}(2), 318--345.

\bibitem[\protect\citeauthoryear{Rosenbaum}{Rosenbaum}{1987}]{rosenbaum1987}
Rosenbaum, P.~R. (1987).
\newblock Sensitivity analysis for certain permutation inferences in matched
  observational studies.
\newblock {\em Biometrika\/}~{\em 74\/}(1), 13--26.

\bibitem[\protect\citeauthoryear{Russell}{Russell}{2020}]{russell2020}
Russell, T.~M. (2020).
\newblock Policy transforms and learning optimal policies.

\bibitem[\protect\citeauthoryear{Stoye}{Stoye}{2009}]{stoye2009}
Stoye, J. (2009).
\newblock Minimax regret treatment choice with finite samples.
\newblock {\em Journal of Econometrics\/}~{\em 151\/}(1), 70--81.

\bibitem[\protect\citeauthoryear{Stoye}{Stoye}{2012}]{STOYE2012138}
Stoye, J. (2012).
\newblock Minimax regret treatment choice with covariates or with limited
  validity of experiments.
\newblock {\em Journal of Econometrics\/}~{\em 166\/}(1), 138--156.
\newblock Annals Issue on ``Identification and Decisions'', in Honor of Chuck
  Manski's 60th Birthday.

\bibitem[\protect\citeauthoryear{Sun}{Sun}{2021}]{sun2021}
Sun, L. (2021).
\newblock Empirical welfare maximization with constraints.

\bibitem[\protect\citeauthoryear{Valiant}{Valiant}{1984}]{valiant}
Valiant, L.~G. (1984).
\newblock A theory of the learnable.
\newblock In {\em Proceedings of the Sixteenth Annual ACM Symposium on Theory
  of Computing}, STOC '84, New York, NY, USA, pp.\  436–445. Association for
  Computing Machinery.

\bibitem[\protect\citeauthoryear{Vapnik}{Vapnik}{1998}]{vapnik1998}
Vapnik, V.~N. (1998).
\newblock {\em Statistical Learning Theory}.
\newblock New York: Wiley.

\bibitem[\protect\citeauthoryear{Viviano}{Viviano}{2019}]{viviano2021}
Viviano, D. (2019).
\newblock Policy targeting under network interference.

\bibitem[\protect\citeauthoryear{Wainwright}{Wainwright}{2019}]{wainwright_2019}
Wainwright, M.~J. (2019).
\newblock {\em High-Dimensional Statistics: A Non-Asymptotic Viewpoint}.
\newblock Cambridge Series in Statistical and Probabilistic Mathematics.
  Cambridge University Press.

\bibitem[\protect\citeauthoryear{Wald}{Wald}{1950}]{wald_1950}
Wald, A. (1950).
\newblock {\em Statistical Decision Functions}.
\newblock Wiley.

\bibitem[\protect\citeauthoryear{Yata}{Yata}{2021}]{yata}
Yata, K. (2021).
\newblock Optimal decision rules under partial identification.

\bibitem[\protect\citeauthoryear{Zhao, Zeng, Rush, and Kosorok}{Zhao
  et~al.}{2012}]{kosorokOWL2012}
Zhao, Y., D.~Zeng, A.~J. Rush, and M.~R. Kosorok (2012).
\newblock Estimating individualized treatment rules using outcome weighted
  learning.
\newblock {\em Journal of the American Statistical Association\/}~{\em
  107\/}(499), 1106--1118.

\end{thebibliography}

\end{document}